\documentclass[format=acmlarge, review=false]{acmart}
\usepackage{acm-ec-21}
\usepackage{booktabs} %
\usepackage[ruled,vlined,linesnumbered]{algorithm2e} %

\SetAlFnt{\small}
\SetAlCapFnt{\small}
\SetAlCapNameFnt{\small}
\SetAlCapHSkip{0pt}
\IncMargin{-\parindent}

\setcitestyle{acmnumeric}

\usepackage{empheq}
\usepackage[capitalize]{cleveref}
\usepackage{textcomp}
\usepackage{bm}
\usepackage{footnote}
\makesavenoteenv{algorithm}

\newtheorem{fact}{Fact}
\crefname{fact}{Fact}{Facts}

\hypersetup{
  colorlinks   = true,
  urlcolor     = teal,
  linkcolor    = teal, 
  citecolor   = teal 
}

\DeclareMathOperator*{\argmax}{\arg\!\max}
\newcommand{\annie}{\emph{Annie}\texttrademark}

\SetKwInput{Parameter}{Parameter}

\newcommand{\Opt}{\textsc{Opt}}

\everypar{\looseness=-1}

\title{Dynamic Placement in Refugee Resettlement}

\colorlet{RED}{red} %
\author{Narges Ahani}
\affiliation{\institution{Worcester Polytechnic Institute}}
\author{Paul G\"olz}
\affiliation{\institution{Carnegie Mellon University}}
\author{Ariel D. Procaccia}
\affiliation{\institution{Harvard University}}
\author{Alexander Teytelboym}
\affiliation{\institution{University of Oxford}}
\author{Andrew C. Trapp}
\affiliation{\institution{Worcester Polytechnic Institute}}

\begin{abstract}
Employment outcomes of resettled refugees depend strongly on where they are placed inside the host country.
Each week, a resettlement agency is assigned a batch of refugees by the United States government.
The agency must place these refugees in its local affiliates, while respecting the affiliates' yearly capacities.
We develop an allocation system that suggests where to place an incoming refugee, in order to improve total employment success.
Our algorithm is based on two-stage stochastic programming and achieves over 98 percent of the hindsight-optimal employment, compared to under 90 percent of current greedy-like approaches. This dramatic improvement persists even when we incorporate a vast array of practical features of the refugee resettlement process including indivisible families, batching, and uncertainty with respect to the number of future arrivals. 
Our algorithm is now part of the \annie\ \textsc{Moore} optimization software used by a leading American refugee resettlement agency.
\end{abstract}

\begin{document}

\maketitle

\section{Introduction}
There are 26 million refugees around the world \citep{unhcr2020global}. The United Nations High Commissioner for Refugees (UNHCR) considers over 1.4 million of them to be in need of \emph{resettlement}, that is, permanent relocation from a temporary country of asylum to the country of resettlement  \citep{unhcr2019projected}. Resettlement is mainly targeted at the most vulnerable refugees, such as children at risk, survivors of violence and torture, and those with urgent medical needs.
Dozens of countries around the world resettle refugees; in 2019, for example, around 63\,000 refugees were resettled~\citep{unhcr2019projected}.
Still, the number of refugees in need of resettlement far exceeds the number that is actually resettled in every year.

Historically, most countries taking in resettled refugees have paid little attention to where inside the country these refugees are placed.
This policy might be worth reconsidering, however, since there is ample evidence that the initial local resettlement destination dramatically affects the outcomes of refugees \citep{aaslund2007and,aaslund2009peer,aaslund2010important,aaslund2011peers,damm2014neighborhood,feywerda2016location,bansak2018improving,marten2019ethnic}.
One specific variable impacted by community placement is whether and when resettled refugees find employment.
Employment plays a key role in the successful integration of a refugee by ``promoting economic independence, planning for the future, meeting members of the host society, providing opportunity to develop language skills, restoring self-esteem and encouraging self-reliance''~\citep{AS08}.

Since promoting employment is so crucial,
the American resettlement agency HIAS began in 2017 to match refugees to communities using the matching software \annie{} \textsc{Moore} (Matching and Outcome Optimization for Refugee Empowerment), which is designed to maximize the total number of refugees who obtain employment soon after arrival~\citep{ahani2020placement}.
Each week, the US goverment assigns a new batch of refugees to HIAS, and \annie{} suggests which community each refugee in the batch should be placed in.
Before this work, \annie{} made its suggestions using a greedy algorithmic approach, that is, each batch of arrivals was allocated by separately maximizing the expected employment of this batch (subject to the remaining community capacities and ensuring that refugees have access to necessary services).
Allocating affiliate capacity in such a greedy way will likely lead to suboptimal employment, however:
A placement algorithm could achieve better employment by weighing in each placement decision whether a slot of capacity is more beneficial when used by a refugee in the current batch or when saved up for some refugee potentially arriving later in the year.

In this paper, we improve the optimization approach of \annie{} by intentionally incorporating the dynamic nature of the matching problem. %
For this, we design two closely related algorithms---one based on stochastic programming and another based on Walrasian equilibrium---that optimize the dynamic matching of refugees to communities in the United States.
Our focus is to study these algorithms in a rich model that captures all of the relevant practical features of the refugee resettlement process, including indivisible families of refugees, batching, and unknown numbers of refugee arrivals. We evaluate the performance of our algorithms on HIAS data from 2014 until 2019. We show that both algorithms achieve over 98 percent of the hindsight-optimal employment in all years whereas the greedy baseline achieves only around 90 percent.
We then describe how we implemented our algorithms within \annie{} to create \annie \ 2.0, which has been well-received by HIAS leadership: ``\annie \ 2.0 is a game-changer for our pre-arrivals processes, allowing us to plan and optimize our pre-arrival strategy a year rather than a week ahead.''

\subsection{Related Work}
Our paper extends a line of work initiated by \citet{bansak2018improving}, which aims to increase refugees' employment outcomes through data- and optimization-driven placement.
This approach consists of two components: using machine learning to estimate the probability that a given refugee placed at a given community would find employment, and using mathematical programming to perform the optimization.
\citet{ahani2020placement} adopted a similar approach to develop \annie; they also pointed out the practical relevance of indivisible families and the possibility of batching. %
Both papers seek to maximize employment with respect to a current batch of refugees, without considering future arrivals; it is in this sense that we think of deployed algorithms as \emph{greedy}, and that is indeed our benchmark in this paper. 

Our dynamic refugee placement problem generalizes the classic \emph{edge-weighted online bipartite matching} problem, but most algorithms in the theoretical literature are not promising for our application since they are optimized for overly pessimistic arrival scenarios.
Whereas competitive analysis was quite successful for unweighted online bipartite matching~\citep{KVV90}, no constant-factor approximation algorithm is possible for the weighted setting if arrivals are adversarial~\citep{FHT+20}.
In the random-order arrival model, a $1/e$-approximation is possible~\citep{KRT+13}, but the algorithm is impractical; in particular, it leaves the first 37\% of arrivals entirely unmatched.
Even if arrivals are drawn i.i.d.\ from a known distribution, \citet{MGS12} show that no online algorithm can obtain a better approximation ratio than 0.823, far below the performance of even the greedy baseline in our setting.
Since this impossibility is based on contrived arrival distributions, many papers additionally assume that arrivals belong to finitely many types determining their edge weights.
In this setting, constructing matchings that are optimal up to lower-order terms (with high probability) is not difficult~\citep[see][]{AHL13}, and multiple papers obtain such results, often in generalizations of edge-weighted online bipartite matching~\citep{AHL12,AHL13,VB21}.
What limits the applicability of these algorithms to our setting, however, is that these algorithms require the distribution over types explicitly in their input, and are often constructed based on the assumption that multiple arrivals of each type will occur in a single run of the algorithm.
By contrast, we estimate employment scores based on 20 independent features, which means the number of refugee ``types'' is too large to enumerate and we do not expect to see identical refugees.

Our algorithmic approach can be seen as an instantiation of the \emph{Bayes Selector}, an algorithmic paradigm that has yielded impressive theoretical and empirical results across various problems with stochastic online arrivals~\citep{FB19,BGV20,VB21,vera2021online,SFC+22}. %
Conceptually, the Bayes Selector takes in a prediction of future arrivals and then performs the action (in our setting: chooses the affiliate for the current arrival) that seems most likely to coincide with the action taken by an optimal benchmark.
Under some regularity conditions on the arrivals, algorithms following this methodology have \emph{constant regret}, that is, the expected difference between the algorithm's performance and that of the optimal benchmark does not grow with the size of the problem.
The prediction of future arrivals often takes other shapes, but can be a sampled trajectory of arrivals as in our work~\citep{BGV20}.
In most papers, the choice of action is based on how \emph{often} the optimal benchmark would take an action in the simulated future rather than, as in our work, on the marginal effect of an action on the optimal value.
Very recently, however, \citet{SFC+22} analyze the same variant of the Bayes Selector (the ``hindsight planning policy'') as our \cref{eq:stochprog}, and show that it gives constant regret for the problem of stochastic online bin packing.
Even though we do not provide theoretical guarantees in this paper, the success of the Bayes Selector across related settings partially explains our good empirical performance.

Shadow prices have been used to guide decisions in online settings in a variety of contexts, including advertising~\citep{VVS10, DH09, vazirani2005adwords}, revenue management~\citep{talluri2004theory}, worker assignment~\citep{ho2012online, johari2021matching}, and resource allocation~\citep{asadpour2020online}.
\cite{agrawal2014dynamic} develop a dynamic learning approach where prices are calculated in a similar manner to ours; while they update their match scores upon every doubling of the arrival history, we update our match scores upon every batch.
\cite{ho2012online} extend the advertising context of~\cite{DH09} to assign workers to tasks when match scores are initially unknown and must be learned.
Like~\cite{ho2012online},~\cite{johari2021matching} also consider the worker-to-job context, but learn scores while matching via an explore-then-exploit approach.
In our setting, our scores are known in advance independent of arrivals~\citep{ahani2020placement}.

In independent and concurrent work, \citet{bansak2020minimum} also considers dynamic refugee resettlement; the algorithm obtaining the highest employment in that study is equivalent to our two-stage stochastic programming formulation in the simplest setting. Our model is much richer as we include non-unit family sizes, incompatibilities between families and communities, and allow for uncertain arrival numbers.
A second limitation of their work is that their best algorithm is prohibitively slow.
This lack of computational efficiency pushes them to consider algorithms with worse employment outcomes, and it limits their empirical evaluation to a single month of arrivals.
By contrast, we leverage multiple algorithmic insights to speed up the algorithm by two orders of magnitude without substantially trading off employment.
This speed-up allows us to empirically evaluate our algorithms for realistic matching periods, which last for an entire fiscal year.
We compare the running times of both algorithms in \cref{sec:containsruntimebansak}.
Finally, very recent work by \citet{bansak2022balancing} extends the earlier \citet{bansak2020minimum} by incorporating a secondary objective that seeks to consume capacity at similar rates across affiliates, improving case wait times across affiliates without sacrificing much employment.
 
\subsection{Organization of the Paper}
In Section~\ref{sec:institutional}, we provide an overview of the US refugee resettlement process. In Section~\ref{sec:model}, we outline our model of dynamic refugee matching. In Section~\ref{sec:unit}, we propose our two algorithms and show that they obtain near-optimal employment %
in a baseline setting that ignores the indivisibility of families, batching, and uncertainty about the total number of arrivals.
In the next three sections, we layer on complexity toward the setting encountered in practice: indivisible families (Section~\ref{sec:nonunit}), batching (Section~\ref{sec:full}), and unknown arrival numbers (Section~\ref{sec:unknownn}).
In these sections, we demonstrate that indivisible families and batching do not substantially change our algorithms' employment performance, and that employment remains high unless the number of arrivals widely deviates from the numbers announced by the government.
In Section~\ref{sec:implementation}, we then explain how we implemented our approach within \annie{} and conclude in Section~\ref{sec:conclusion}. In the appendix, we provide deferred details and additional empirical analyses.

\section{Institutional Background}\label{sec:institutional}
The federal Office of Refugee Resettlement was created by the Refugees Act in 1980. The Act established funding rules and authorized the President of the United States to set annual capacities for resettlement. The resettlement process is managed by the US Refugee Admissions Program (USRAP) of the US Department of State, in conjunction with a number of federal agencies across federal departments as well as the International Organization for Migration and the UNHCR.

Applications for the resettlement program take place from outside of the US, typically in refugee camps. The US government conducts security checks, medical screening, and performs cultural orientation, which can take upwards of 18 months~\citep{Jones15}.
After clearance, USRAP decentralizes the process of welcoming refugees to nine NGOs known as \emph{resettlement agencies}, of which one is HIAS.
Each agency works with their own network of local \emph{affiliates}, each supported by local offices as well as religious entities like churches, synagogues, or mosques, which serve as community liaisons for refugees.
Each agency typically works with dozens of affiliates, though the number of affiliates can fluctuate over time. Some affiliates lack services to host certain kinds of refugees. For example, certain affiliates do not have translators for non-English-speaking refugees or lack support for single-parent families.

Agencies have no influence on which refugees are cleared for resettlement by USRAP or on when the refugees might arrive.  Resettlement agencies meet on a weekly or fortnightly basis in order to allocate among themselves the refugees that have been cleared by USRAP.
Refugees are usually resettled with members of their family.
Such an indivisible group of refugees is referred to as a \emph{case}.
As a family can split when its members are fleeing their home country, some refugees who are applying for resettlement might already have existing relatives or connections in the US. Such cases \emph{with US ties} are automatically resettled near their existing ties.
All other refugees, referred to as \emph{free cases}, can be resettled by any agency into any of the agency's affiliates.

Each affiliate has an assigned annual capacity for the number of refugees (rather than cases) it can admit in a given fiscal year.\footnote{Each fiscal year ranges from October 1 of the previous calendar year to September 30. For example, fiscal year 2017 ranges from October 1, 2016 to September 30, 2017.} These capacities are approved by USRAP and, in theory, agencies cannot exceed them. In practice, capacities can be slightly adjusted towards the end of the year or, as in recent years, substantially revised in the course of the year.
Since capacities limit the number of refugees \emph{arriving} in a fiscal year rather than \emph{allocated} in it, and since there is typically a delay of multiple months between the two events, the Department of State tells the resettlement agencies an estimated arrival date for each cleared case.
Agencies are assessed annually by USRAP on their performance in finding employment for refugees within 90 days of their arrival. Data on 90-day employment is therefore diligently collected by the affiliates and monitored by the agencies.

\section{Model}\label{sec:model}
An instance of the matching problem first defines a set $L$ of \emph{affiliates}, and each affiliate $\ell$ has a capacity $c_\ell \in \mathbb{N}_{\geq 0} \cup \{\infty\}$ of how many refugees it can host.
We call a collection $\{c_\ell\}_{\ell \in L}$ of capacities for all affiliates a \emph{capacity profile} $\bm{c}$.
To describe changes in capacity, it will be useful to manipulate the capacity profiles as vectors.
Specifically, we write $\bm{c} - \bm{e}_\ell$ to describe the capacity profile obtained from $\bm{c}$ by reducing the capacity of affiliate $\ell$ by 1.

On the other side of the matching problem is a set $N = \{1, \dots, n\}$ of \emph{cases}.
Each case $i$ represents an indivisible family of $s_i \in \mathbb{N}_{\geq 1}$ refugees.
Furthermore, each case $i$, for each affiliate $\ell$, has an \emph{employment score} $u_{i, \ell}$, which indicates the expected number of case members that will find employment if the case is allocated to $\ell$.
Typically, these employment scores $u_{i, \ell}$ are real numbers in $[0, s_i]$, but we will also allow to set $u_{i, \ell} = - \infty$ to express that case $i$ is not compatible with affiliate $\ell$.
We will refer to the combination of a case's size and vector of employment scores as the \emph{characteristics} of the case. To ensure that the matching problem is always feasible, we will assume that $L$ contains a special affiliate $\bot$ that represents leaving a case unmatched, where $u_{i, \bot} = 0$ for all cases $i$ and $c_\bot = \infty$.\footnote{For example, allowing cases to be unmatched is necessary since an arriving case might only be compatible with affiliates whose capacity is already exhausted. When these situations occur in practice, such cases do not remain unmatched; instead, capacities can be increased or case--affiliate incompatibilities overruled manually by the arrivals officer. For our sequence of models, we report the fraction of matched refugees in \cref{sec:unmatched}, and find that our algorithms do not lead to fewer refugees being matched than in the greedy baseline. To lower the number of unmatched refugees at the cost of reducing employment, one can add a constant reward per refugee to the $u_{i, \ell}$ with $\ell \neq \bot$.}

We use the employment scores developed by \citet{ahani2020placement}, and we give details on data preprocessing and training in \cref{app:datacleaning}.
Throughout this paper, we consider these employment scores as ground truth, which means that we evaluate algorithms directly based on the employment scores.
An evaluation of how accurately the employment scores predict employment outcomes is outside of the scope of this paper, and has already been performed by Ahani et al.

The goal of the matching problem is to allocate cases to affiliates such that the \emph{total employment}, that is, the sum of employment scores, is maximized, subject to not exceeding capacities.
For a set $I \subseteq N$ and a capacity profile $\bm{c} = \{c_\ell\}_{\ell \in L}$, define $\textsc{Matching}(I, \bm{c})$ as the matching integer linear program (ILP) below, where variables $x_{i, \ell}$ indicate whether case $i \in I$ is matched to affiliate $\ell \in L$:
\begin{align*}
    \text{maximize}~& \sum_{i \in I} \sum_{\ell \in L} u_{i, \ell} \, x_{i, \ell} & \\
    \text{subject to}~& \sum_{\ell \in L} x_{i, \ell} = 1 & \forall i \in I \\
    & \sum_{i \in I} s_i \, x_{i, \ell} \leq c_\ell & \forall \ell \in L \\
    & x_{i, \ell} \in \{0, 1\} & \forall i\in I, \ell \in L.
\end{align*}
Let $\Opt(I, \bm{c})$ denote the optimal objective value of $\textsc{Matching}(I, \bm{c})$.
The \emph{linear programming} (LP) \emph{relaxation} of $\textsc{Matching}(I, \bm{c})$ is obtained by replacing the constraint $x_{i, \ell} \in \{0, 1\}$ by $0 \leq x_{i, \ell} \leq 1$ for all $i\in I, \ell \in L$.
For a fixed matching, we define the \emph{match score} of a case $i$ as its employment score $u_{i, \ell_i}$ at the affiliate $\ell_i$ where it is allocated; we will also refer to its \emph{match score per refugee}, $u_{i, \ell_i} / s_i$.

Finally, cases arrive \emph{online}, that is, they arrive one by one and, when case $i$ arrives, the decision of which affiliate to place it in must be made irrevocably, before the characteristics of the subsequent arrivals $i+1, \dots, n$ are known.\footnote{From \cref{sec:full} onward, cases will instead arrive in batches, which can be allocated simultaneously.}
Thus, although an online matching algorithm must still produce a matching whose indicator variables $x_{i, \ell}$ satisfy the constraints of $\textsc{Matching}(N, \bm{c})$, the total employment $\sum_{i \in N, \ell \in L} u_{i, \ell} \, x_{i, \ell}$ typically will not attain the benchmark $\Opt(N, \bm{c})$ of the \emph{optimal matching in hindsight}.
While we will not commit to a specific model of how the characteristics of arriving cases are generated, these arrivals should be thought of as \emph{stochastic} rather than worst-case, and the distribution of case characteristics as changing slowly enough that sampling from recent arrivals is a reasonable proxy for the distribution of future arrivals.

Throughout the following sections, we will consider a sequence of models, which incorporate an increasing number of features of the real-world refugee allocation problem: in \cref{sec:unit}, we consider traditional online bipartite matching, which results from requiring $s_i = 1$ in the above model; from \cref{sec:nonunit} onward, we allow cases to have arbitrary size; from \cref{sec:full} onward, we also allow cases to arrive in batches rather than one by one; in \cref{sec:unknownn}, we no longer assume that the total number $n$ of arriving cases is known to the algorithm.

\section{Online Bipartite Matching ($s_i=1$)}
\label{sec:unit}

In this section, we will consider the special case of online bipartite (weighted) matching.
We stress that this classic problem does not capture key features of the refugee-allocation problem in practice, which we will add in later sections.
Instead, online bipartite matching allows us to more cleanly draw connections to theoretical arguments, which help motivate our algorithm design.
Later in the paper, we will empirically show that the approach continues to work well in richer and more realistic settings.

Formally, this section considers the model defined in the previous section, with the restriction that all cases consist of single refugees, that is, that $s_i = 1$ for all $i \in N$.
Under this assumption, it is well-known that the optimum matching for the ILP $\textsc{Matching}(I, \bm{c})$ can be found by solving its LP relaxation.

\subsection{Algorithmic Approach}
\label{sec:potentialapproach}
To motivate our algorithmic approach, we begin by describing why matching systems currently deployed in practice lead to suboptimal employment.
These systems assign cases \emph{greedily}, which---putting aside batching for now---means that an arriving case $i$ is matched to the affiliate $\ell$ with highest employment score $u_{i, \ell}$ among those that have at least $s_i$ remaining capacity.
The main problem with greedy assignment is that it exhausts the capacity of the most desirable affiliates too early.
In particular, we observe on the real data that a large fraction of cases have their highest employment score in the same affiliate $\ell^*$, but that the size of the employment advantage of affiliate $\ell^*$ over the second-best affiliate varies.
Since it only considers the highest-employment affiliate for each case, greedy assignment will fill the entire capacity of $\ell^*$ early in the year, including with some cases that benefit little from this assignment.
Consequently, cases that would particularly profit from being placed in $\ell^*$ but arrive later in the year no longer fit within the capacity.

Intuitively, the decision to match a case $i$ to an affiliate $\ell$ has two effects: the immediate increase of the total employment by $u_{i, \ell}$ but also an opportunity cost for consuming $\ell$'s capacity, which might prevent profitable assignments for later arrivals.
Since greedy assignment only considers the former effect, it leaves employment on the table.

A better approach is \emph{two-stage stochastic programming}, which allocates an arriving case $i$ to the affiliate $\ell$ maximizing the sum of the immediate employment $u_{i, \ell}$ and the expected optimal employment obtainable by matching the future arrivals subject to the remaining capacity.
That is, if, at the time of $i$'s arrival, the remaining capacities are given by $\bm{c}$, two-stage stochastic programming allocates $i$ to the affiliate
\begin{align}
&\argmax_{\ell \in L : c_\ell \geq s_i} \quad u_{i, \ell} + \mathbb{E} \Big[\Opt\big(\{i+1, \dots, n\}, \bm{c} - s_i \cdot \bm{e}_{\ell}\big)\Big], \notag \\
\intertext{where the expectation is taken over the characteristics of cases $j = i+1, \dots, n$. Since adding a constant term does not change the argmax, this can be rewritten as}
= &\argmax_{\ell \in L : c_\ell \geq s_i} \quad u_{i, \ell} - \mathbb{E} \Big[\Opt\big(\{i+1, \dots, n\}, \bm{c}\big)\Big] + \mathbb{E} \Big[\Opt\big(\{i+1, \dots, n\}, \bm{c} - s_i \cdot \bm{e}_\ell \big)\Big] \notag \\
= &\argmax_{\ell \in L : c_\ell \geq s_i} \quad u_{i, \ell} - \mathbb{E} \Big[\Opt\big(\{i+1, \dots, n\}, \bm{c}\big) - \Opt\big(\{i+1, \dots, n\}, \bm{c} - s_i \cdot \bm{e}_\ell \big)\Big]. \label{eq:stochprog}
\intertext{Using our assumption that $s_i = 1$, this can be simplified to}
= &\argmax_{\ell \in L : c_\ell \geq 1} \quad u_{i, \ell} - \mathbb{E} \Big[\Opt\big(\{i+1, \dots, n\}, \bm{c}\big) - \Opt\big(\{i+1, \dots, n\}, \bm{c} - \bm{e}_\ell \big)\Big].\notag
\end{align}
Note that the expectation that is subtracted in either of the last two lines is exactly the expected opportunity cost of reducing the capacity of $\ell$ by placing case $i$ there.
This motivates our algorithmic approach:
in every time step, we first compute a \emph{potential} $p_\ell$ for each affiliate $\ell$.
Then, rather than myopically maximizing the utility of the match as does greedy assignment, our algorithm \textsf{PM} (``\textbf{p}otential \textbf{m}atch'') myopically maximizes the utility of the current match minus the potential of the capacity used, as shown in \cref{alg:nobatching}.
(Note that an affiliate $\ell$ can always be defined in \cref{lin:ellnobatching} as, by assumption, $c_\bot = \infty$.)

\begin{algorithm}[htb]
\Parameter{a subroutine $\textsf{Potential}$ to determine affiliate potentials}
initialize the capacities $c_\ell$ for each affiliate $\ell$\;
\For{$t = 1, \dots, n$}{
observe the case size $s_t$ and the employment scores $\{u_{t, \ell}\}_{\ell}$\;
call $\textsf{Potential}()$ to define a potential $p_{\ell}$ for each affiliate $\ell$\;
    $\ell \leftarrow \argmax_{\ell \in L : c_\ell \geq s_t} u_{t, \ell} - s_t \, p_\ell$\; \label{lin:ellnobatching}
    allocate case $t$ to $\ell$ and set $c_\ell \leftarrow c_\ell - s_t$\;
}
\caption{\textsf{PM}(\textsf{Potential})}
\label{alg:nobatching}
\end{algorithm}

We estimate the expected value of the opportunity cost by averaging over a fixed number $k$ of \emph{trajectories}, each of which consists of randomly sampled characteristics of all arrivals $i+1$ through $n$.
As the characteristics of arriving refugees change over time, and as these changes tend to be gradual, we draw these arrival characteristics uniformly with replacement from the arrivals in the six months prior to the current allocation decision.
In \cref{app:lookback}, we evaluate different lengths of this sampling window.

For each sampled trajectory, it remains to calculate the potential, which we would like to equal the opportunity cost $\Opt\big(\{i+1, \dots, n\}, \bm{c}\big) - \Opt\big(\{i+1, \dots, n\}, \bm{c} - \bm{e}_\ell\big)$.
Clearly, this could be computed by solving $\mathcal{O}(|L|)$ matching LPs, which is what the flagship algorithm by \citet{bansak2020minimum} does.

Instead, we make use of a celebrated result in matching theory~\citep{Leonard83} to compute the opportunity costs for \emph{all} affiliates with remaining capacity as the shadow prices of a \emph{single} LP:
\begin{fact}
Fix a matching-problem instance, in which all cases $i$ have size $s_i = 1$.
In the LP relaxation of $\textsc{Matching}(N, \bm{c})$, let $\{p_\ell\}_{\ell \in L}$ denote the unique element-wise maximal set of shadow prices for the constraints $\sum_{i \in N} s_i \, x_{i, \ell} \leq c_\ell$.
Then, for each $\ell$ with $c_\ell \geq 1$,
\[ p_\ell = \Opt\big(\{i+1, \dots, n\}, \bm{c}\big) - \Opt\big(\{i+1, \dots, n\}, \bm{c} - \bm{e}_\ell \big).\]
\end{fact}
This suggests the procedure $\textsf{Pot1}$ for computing potentials, which is shown in \cref{alg:potential1}.

We also develop a second method \textsf{Pot2} for computing potentials, which is based on a slightly different LP and has different theoretical underpinnings:
\begin{itemize}
    \item whereas the matching LP for \textsf{Pot1} does not include the current batch of arrivals, the current batch is included in the LP for \textsf{Pot2},
    \item whereas \textsf{Pot1} uses the element-wise maximal set of shadow prices, \textsf{Pot2} uses the element-wise minimal one, and
    \item whereas \textsf{Pot1} is theoretically derived from two-stage stochastic programming, \textsf{Pot2} is motivated by a connection to Walrasian equilibria.
\end{itemize}
For conciseness, we defer the formal definition of \textsf{Pot2} and its connection to the Walrasian equilibrium to \cref{app:potential2}.

\subsection{Empirical Evaluation}
\label{sec:onlinebipartiteevaluation}

We evaluate the employment of our potential-based matching algorithm on real yearly arrivals at HIAS.
For each fiscal year, we consider all refugees who arrived in this period, and we consider them in the order in which they were received for allocation by HIAS.
For the capacities, we use the year's \emph{final}, i.e. most revised, capacities.\footnote{When the number of refugees resettled in the fiscal year exceeds the official capacity, we use the number of resettled refugees instead. In these situations, HIAS negotiated an increase in capacity that is not always recorded in our data.}
We also immediately take into account that affiliates are restricted in which nationalities, languages, and family sizes they can accommodate, as well as in whether they can host single parents and the constraints on tied cases.

\begin{algorithm}
\Parameter{$k \in \mathbb{N}_{\geq 1}$, the number of trajectories per potential computation}
\KwIn{remaining capacities $\bm{c}$, the index $t$ of the last observed case, characteristics of cases arriving in the past 6 months}
\KwOut{a set of potentials $p_\ell$ for all affiliates $\ell$}

\For{$j = 1, \dots, k$}{
for each $i = t+1, \dots, n$, set $s_{i}$ and $\{u_{i, \ell}\}_{\ell}$ to the size and employment scores of a random, recently arrived case\;
solve the following bipartite-matching LP:
\begin{empheq}[box=\fbox]{align*}
    \text{maximize}~& \sum_{i=t+1}^n \sum_{\ell \in L} u_{i, \ell} \, x_{i, \ell} \\
    \text{subject to}~& \sum_{\ell \in L} x_{i, \ell} = 1 & \forall i = (t\!+\!1), \dots, n \\
    & \sum_{i = t+1}^n s_i \, x_{i, \ell} \leq c_{\ell} & \forall \ell \in L & \quad (*) \hspace{3mm} \label{eq:algcapacity} \\
    & 0 \leq x_{i, \ell} & \forall i = (t\!+\!1), \dots, n, \forall \ell \in L.
\end{empheq}\\
for each $\ell$, set $p_{\ell}^j$ to be the maximal shadow price\footnotemark{} %
of the constraint $(*)$\;
}
set $p_\ell \leftarrow (\sum_{j=1}^k p_{\ell}^j) / k$ for all $\ell$\;
\Return $\{p_\ell\}_{\ell \in L}$\;
\caption{\textsf{Pot1}($k$)}
\label{alg:potential1}
\end{algorithm}
\footnotetext{One way of finding the maximal shadow price is to first solve the dual LP to find its objective value, then adding a constraint that constrains the objective of the dual LP to be equal to this optimal objective value, and to finally maximize the sum of dual variables $p_\ell$ over this new restricted LP.}

The main way in which this experiment deviates from reality is the assumption (made throughout this section) that cases have unit size.
To satisfy this assumption, we split each case of size $s_i > 1$ into $s_i$ identical single-refugee cases with a $1/s_i$ fraction of the original employment scores.
In subsequent sections, we will repeat the experiments without this modification.

We study 6 fiscal years, from 2014 to 2019.
As affiliates closed and opened across these years, the number of affiliates varies between 16 and 24 (not counting the unmatched affiliate $\bot$).
Finally, the number of arriving refugees (respectively, cases) varies between 1\,670 (resp., 640) and 4\,150 (resp., 1\,630) across fiscal years.

\begin{figure}[tbh]
\centering
\includegraphics[width=\textwidth]{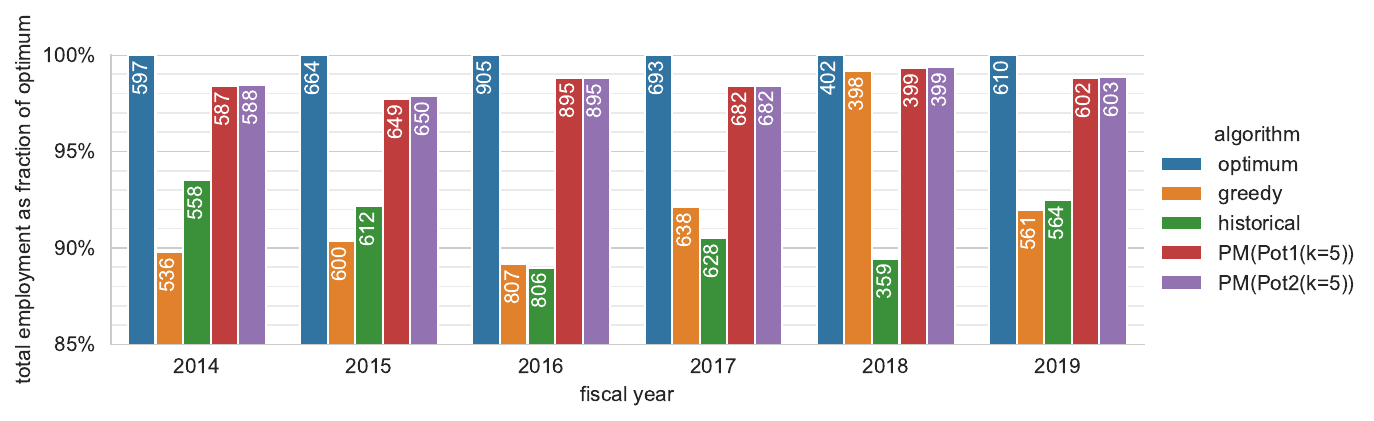}
\caption{Total employment obtained by different algorithms, assuming that cases are split into multiple cases of size $1$. Capacities are the final capacities of the fiscal year. For the potential algorithms, total employment is averaged over 10 random runs. The numbers in the bars denote the absolute total employment; the bar height indicates the proportion of the optimum total employment in hindsight.}
\label{fig:unit}
\end{figure}

As shown in \cref{fig:unit}, even the greedy baseline obtains a total employment of between 89\% and 92\% of $\Opt(N, \bm{c})$, the optimum matching in hindsight.
(One outlier is the year 2018, which we discuss below.)
Nevertheless, the greedy algorithm leads to between 50 and 100 fewer refugees finding employment every year compared to what would have been possible in the optimum matching.
Our potential algorithms close a large fraction of this gap, obtaining between 98\% and 99\% of the optimal total employment, both for algorithms based on \textsf{Pot1} and for those based on \textsf{Pot2}.
Since experiments in this model take much longer to run than those in subsequent models, we defer a comparison between the two potential methods and between values of $k$ to \cref{sec:fullempirics}, where we can run the potential algorithms a sufficient number of times to discern smaller differences.

The fiscal year 2018 stands out from the others due to the fact that the greedy algorithm performs on par with the potential algorithms, at 99\% of the hindsight-optimal total employment.
This is easily explained by the fact that the capacities are much looser than in other fiscal years:
whereas, in all other fiscal years between 2014 and 2019, the number of arriving refugees amounts to between 84\% (2019) and 97\% (2016) of the final total capacity across all affiliates, this fraction is only 48\% in 2018.
Since capacity is so abundant, the optimal matching will match a large fraction of cases to their maximum-score affiliate, and the greedy matching is close to optimal.

We also compare to the employment obtained by the allocation chosen by HIAS (``historical'').
This comparison gives the historical matching a slight advantage, as HIAS sometimes overrides the incompatibility between an affiliate and a case, which we do not allow any other algorithm to do.\footnote{In these cases, we estimate the employment achieved by the case using the regression rather than using $u_{i, \ell} = - \infty$.}

\begin{figure}[tbh]
\centering
\includegraphics[width=\textwidth]{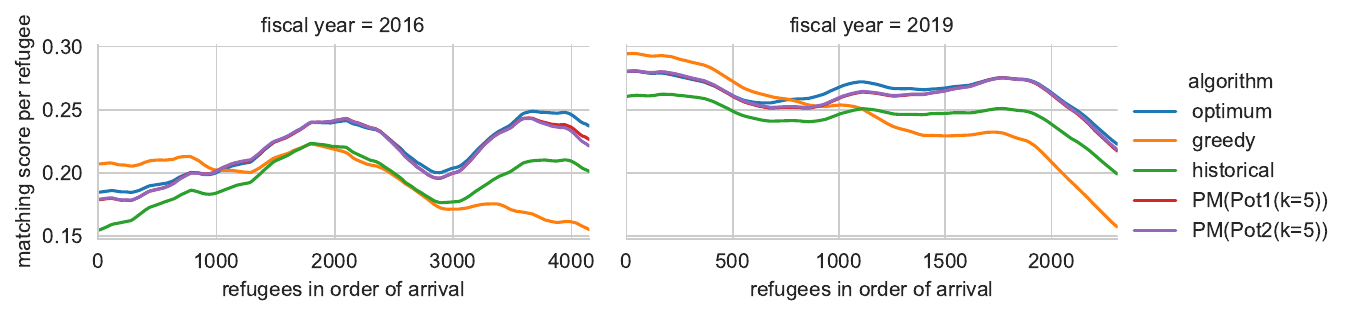}
\caption{Evolution of the per-refugee match score in order of arrival, for fiscal years 2016 and 2019 in the experiment of \cref{fig:unit} (split cases, final capacities). Consecutive match scores are smoothed using triangle smoothing with width 500.}
\label{fig:smoothedunit}
\end{figure}
\looseness=-1
In \cref{fig:smoothedunit}, we investigate how the match score changes over the course of two fiscal years, 2016 and 2019, chosen to contain one year in which the greedy and historical baselines perform relatively poorly (2016) and one in which they perform well (2019).
As the match score of subsequently arriving refugees can greatly differ, these graphs are heavily smoothed over time.
If arrivals were drawn from a time-invariant distribution, we would expect the curves for the optimum matching in hindsight to be level, since how much employment the optimum matching can extract from a case would be independent of the case's arrival time.
Instead, we see that the employment prospects of arrivals fluctuate noticeably over time; in particular, the early refugees in fiscal year 2016 and the late refugees in fiscal year 2019 seem to have worse employment prospects than other refugees in the plot.

The curves for both potential algorithms are nearly indistinguishable from one another, which shows that the algorithms make very similar decisions.
In 2016, these curves start out closely tracking the curve of the optimal-hindsight matching, but fall behind for the last cases, which we observe in most fiscal years.
The similarity of the curves over most of the year indicates that our approach of sampling trajectories from past arrivals is nearly as useful as the optimum algorithm's perfect knowledge of future arrivals and that it leads to a similar trade-off in extracting immediate employment versus preserving capacity for later arrivals.
Of course, the imperfect knowledge of the future incurs a small loss towards the end of the fiscal year, likely because the amount of capacity reserved per affiliate does not perfectly match the demand, which explains the gap in total employment between the hindsight optimum and the potential algorithms.
This typical end-of-year effect is not very pronounced in fiscal year 2019, likely because the final arrivals of fiscal year 2019 have lower employment probabilities than what would be expected based on past arrivals.
Instead, the potential algorithms fall behind the optimum algorithm for some period in the middle of the year, perhaps because they are reserving capacity for late arrivals which the optimum already knows to hold little promise.

\begin{figure}[tb]
\centering
\includegraphics[width=\textwidth]{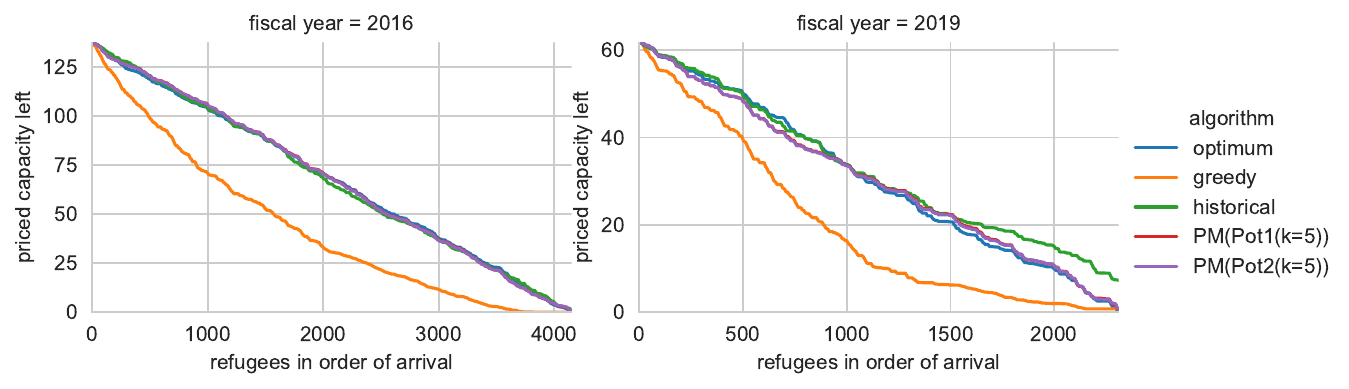}
\caption{Remaining priced capacity at the time of arrival of different refugees, for fiscal years 2016 and 2019 in the experiment of \cref{fig:unit} (split cases, final capacities).}
\label{fig:pricedcapsunit}
\end{figure}
The most striking curve is that of the greedy algorithm, which lies above those of all other algorithms in the first quarter of arrivals, but then falls clearly below the other curves in the second half.
This observation can be explained by the effect we predicted in the motivation of our potential approach: the greedy algorithm extracts small additional gains in employment early in the arrival period, at the cost of prematurely consuming the capacity of the most desirable affiliates.
Then, the lack of capacity limits the match scores of later arrivals, resulting in an overall unfavorable trade-off.
This effect can be directly seen in \cref{fig:pricedcapsunit}, in which we visualize the amount of capacity remaining in the most valuable affiliates.
Specifically, looking at all arrivals of the fiscal year, we compute the shadow prices of the matching LP.
At any point in time, we can then weight the remaining capacity  by these prices to obtain a \emph{priced capacity}.
In \cref{fig:pricedcapsunit}, we see that the optimum-hindsight matching and the potential algorithms use up the priced capacity at a roughly constant pace and essentially consume it all.
By contrast, the greedy algorithm uses up the capacity very quickly, such that at the median refugee, only 22\% (2016) or 17\% (2019) of the priced capacity is left.

The historical matching made by HIAS does not have such obvious defects, but still falls short in terms of total employment.
In both reference years, the average employment moves in parallel with the optimum matching, meaning that HIAS does not overly focus on extracting employment at certain parts of the fiscal year at the expense of others.
However, the average employment consistently lies below that of the optimum and of the potential algorithms.
We see in \cref{fig:pricedcapsunit} that, in 2019, HIAS started consuming the priced capacity at a near-constant pace very similar to that of the optimum algorithm.
Around the median arrival, however, the historical matching slowed down its capacity consumption and ended up not consuming all priced capacity, which explains some loss in total employment.
One reason for this behavior might be that HIAS staff treat the last 9\% of the capacity as a reserve that they are more reluctant to use.
In a year such as 2019, in which the overall arrivals were only 84\% of the total capacity, this heuristic might have actually kept much of the reserve capacity free, including in the affiliates that could have generated higher employment.
By contrast, the total arrivals in 2016 amounted to 97\% of the overall capacity, which could explain why nearly all priced capacity was consumed in this year.
Despite using up priced capacity in a similar pattern as the optimum matching in 2016, the historical assignment achieved lower matching scores throughout the year.
This indicates that the low employment of the historical matching is not just due to a reluctance to use the entire capacity, but that the priced capacity is furthermore inefficiently allocated.

\section{Non-Unit Cases ($s_i \geq 1$)}
\label{sec:nonunit}

The most pressing aspect of refugee matching that we have ignored thus far is that many cases do not consist of individual refugees.
Instead, they consist of an entire family of refugees, which has to be resettled to the same affiliate.

To accommodate cases consisting of multiple family members, we will from now drop the assumption that the $s_i$ are 1.
The main effect of this change is that the LP relaxation of the ILPs $\textsc{Matching}(I, \bm{c})$ can now be a strict relaxation.
Indeed, the LP relaxation might allow for higher objective values because it allows fractional solutions.\footnote{One can always find a fractional solution that splits cases into $1/s_i$ fractions similarly to what we did in the evaluation of \cref{sec:onlinebipartiteevaluation}.}
As a result, our dual prices will no longer \emph{exactly} compute the marginal value of a unit of capacity.
In any case, to retain the exact connection to stochastic programming in \cref{eq:stochprog}, \textsf{PM} would have to subtract the opportunity cost of $s_i$ units of capacity from $u_{i, \ell}$, which might exceed $s_i$ times the opportunity cost of a single unit of capacity.

However, as the capacity of most affiliates is much larger than the size of a typical case, both approximations can be expected to be relatively close, which is what we find empirically:
we repeat the experiment of the previous section, but without splitting up cases into individual refugees.
The results are nearly indistinguishable, which supports our decision to use LP relaxations even in the setting with indivisible cases.
The full figures are deferred to \cref{app:figs:nobatching}.

\section{Batching}
\label{sec:full}
A second aspect that we have not considered thus far is that HIAS does not actually process arriving cases one by one, but in batches containing one or multiple cases.
Most of these batches result from the weekly meetings between the resettlement agencies, but smaller batches with urgent cases are allocated between the weekly meetings.

The fact that cases arrive in batches does not make the problem harder; after all, a matching algorithm that does not support batching can still be used by presenting the cases of each batch to the algorithm one by one.
As we will argue, however, batching represents an opportunity to improve on this strategy: there is a (limited) opportunity to increase total employment and a (substantial) opportunity to reduce running time.

Concerning total employment, using a non-batching algorithm in a batching setting is wasteful since it ignores potentially valuable information.
Specifically, when the earliest cases of the batch are allocated, a non-batching algorithm presumes that the characteristics of the other cases in the batch are not yet known.
Arguably, as the sizes of batches tend to be much smaller than the total number of cases $n$, the amount by which accounting for this information can increase total employment is likely to be limited.

As for running time, given that the matching algorithm receives no new information between the first and last case of a batch, it seems reasonable not to recompute potentials within a batch.
As there tend to be 5 to 10 times more cases than batches and as the computation of potentials is the bottleneck in the running time of the potential algorithms, this promises to substantially speed up the algorithm.

In adapting our algorithm \textsf{PM} to batching, we will not change how we compute the potentials $p_\ell$.
However, the algorithm now allocates all cases in the batch at once, still with the objective of optimizing the immediate utility of the assignment less the sum of potentials consumed.
Thus, our extended algorithm \textsf{PMB} (``\textbf{p}otential \textbf{m}atch with \textbf{b}atching'', \cref{alg:batching} in \cref{app:alg:batching}) allocates the current batch according to the solution to a matching ILP, in which the utility of matching case $i$ to affiliate $\ell$ is set to $u_{i, \ell} - s_i \, p_\ell$.
Note that, if all batches have size $b=1$, this algorithms coincides with our previous algorithm \textsf{PM}.
Moreover, \textsf{PMB} also generalizes the greedy algorithm previously implemented in \annie, which can be recovered by setting all potentials $p_\ell$ to zero.

\label{sec:containsruntimebansak}We can now compare the running time of our algorithms to the flagship algorithm by \citet{bansak2020minimum}, which is very closely related, but does not use dual prices to compute opportunity costs and handles batching in a way that does not improve running time.
The computational bottleneck in both of our algorithms and theirs is the computation of bipartite-matching LPs over the trajectories of simulated future arrivals.
Whereas we compute a single such program per batch of arrivals, Bansak solves $|L| \cdot b$ many such LPs per batch, where $|L|$ is the number of affiliates and $b$ is the number of cases in the batch.
In our dataset, a typical value of $|L| \cdot b$ is around 150, so these speed-ups are substantial.

\subsection{Empirical Evaluation}
\label{sec:fullempirics}
\begin{figure}[tbh]
\centering
\includegraphics[width=\textwidth]{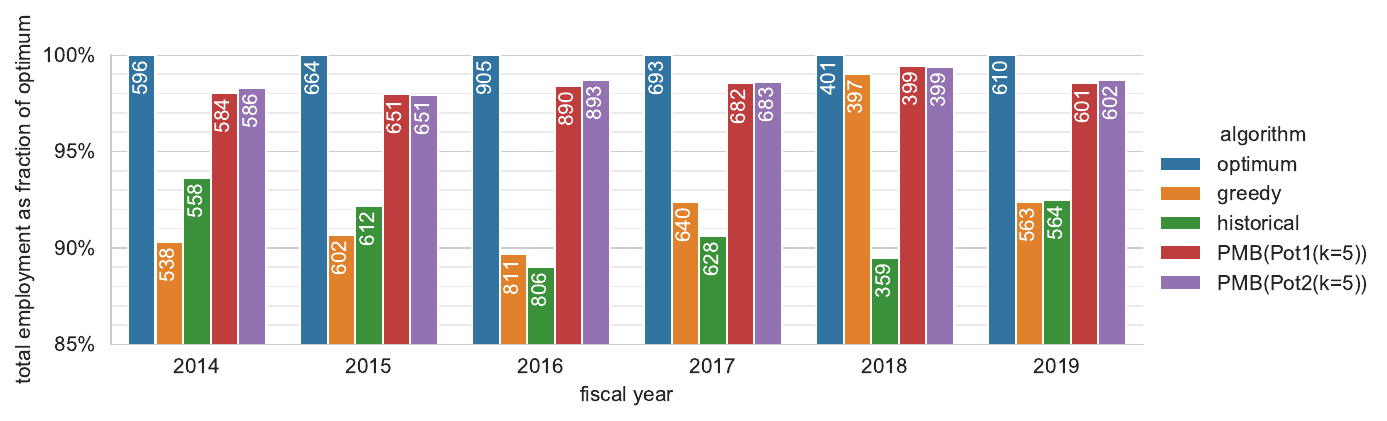}
\caption{Total employment, where cases are not split and arrive in batches. Capacities are the final fiscal year capacities. In contrast to \cref{fig:unit}, cases are treated as indivisible, cases arrive in batches, and the batching variants of greedy and the potential algorithms are used.
For the potential algorithms, the mean employment across 50 random runs is shown.}
\label{fig:full}
\end{figure}
We repeat the experiment measuring the total employment obtained by the algorithms, this time with the greedy algorithm and the potential algorithms allocating cases in batches.
As shown in \cref{fig:full}, the results again look very close to those in the restricted setting of online bipartite matching, confirming that our algorithmic approach generalizes well not only to non-unit case sizes but also to batching as it is used in practice.

\begin{figure}
    \centering
    \includegraphics[width=\textwidth]{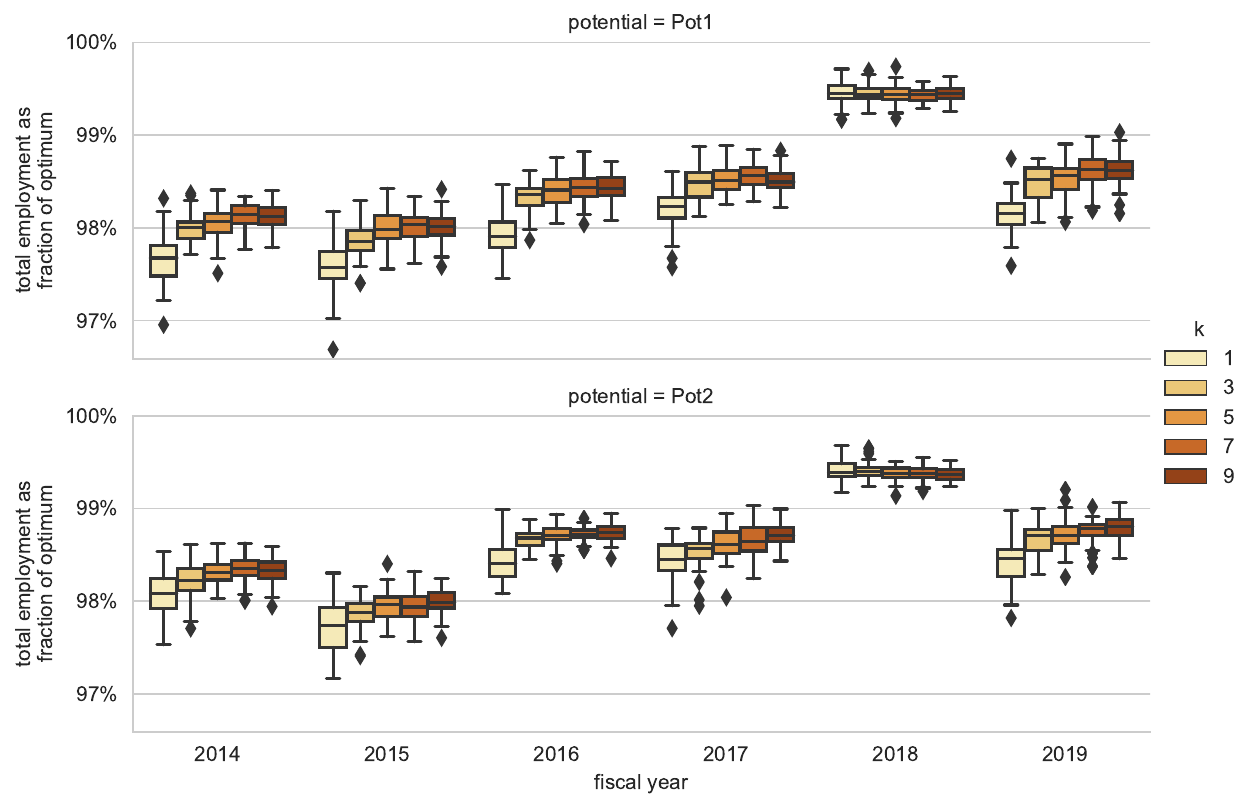}
    \caption{Distribution of the total employment obtained by instantiating \textsf{PMB} with different potential methods and different $k$, in the experiment of \cref{fig:full} (whole cases, batching, final capacities) and over 50 random runs per algorithm.}
    \label{fig:boxesfull}
\end{figure}
Since processing entire cases in batches is much faster than processing cases (or individual refugees) one by one, we are now in a position to run each potential algorithm many times and analyze the distribution of total employments.
As shown in \cref{fig:boxesfull}, the total employment produced by each potential algorithm is sharply concentrated, especially when the algorithms use $k \geq 3$ trajectories to compute duals.

Running each algorithm many times enables us to compare the relative performance of the potential algorithms.
Across both ways of computing potentials, and all fiscal years (with the exception of 2018, where everything is very close together), we see a clear tendency that averaging the potentials across more trajectories improves the employment outcome.
These effects are somewhat limited, though, as going from a single trajectory to nine trajectories improves the median employment by less than half a percent of the hindsight optimum.
As is to be expected, increasing $k$ exhibits diminishing returns.

For $k$ held constant, we observe that the \textsf{Pot2} variants quite consistently outperform the \textsf{Pot1} variants; again with the exception of 2018, in which a small inversion of this trend can be seen.
While all potential algorithms perform very well, based on these results, we recommend the \textsf{Pot2} potentials with a relatively large $k$ for practical implementation.
Of course, increasing $k$ increases the running time of the matching algorithm.
However, since a resettlement agency computes only one set of potentials per day, the algorithm runs in few seconds even for $k=9$ (see \cref{app:timing}).

To additionally support our observation that the potential algorithms outperform the greedy algorithm and the historical matching, we repeat the experiment from \cref{fig:full} for additional arrival sequences derived from the historical data.
As we show in \cref{app:reversedetc}, we obtain similar employment performance as in \cref{fig:full} if the arrival sequence for each year is reversed, or if we consider shifted yearly arrival periods from, say, April to the March of the following year rather than fiscal years (from October to September).
In \cref{sec:bootstrap}, we also evaluate the algorithms on bootstrapped arrivals.
While we discuss more specific observations there, the potential algorithms perform similarly well or slightly better in that setting, consistently at 99\% of the hindsight optimum.

\section{Uncertainty in the Number of Future Arrivals}
\label{sec:unknownn}
Given that our algorithm \textsf{PMB} supports non-unit sized cases and batching, it might seem that we are ready to replace the greedy algorithm in \annie\ by our potential algorithm.
However, our algorithm crucially relies on one piece of input that the greedy algorithm did not need, namely, the total number of cases arriving in the fiscal year.
This number determines the length of the sampled trajectories, which can greatly impact the shadow prices and, thus, how the algorithm allocates cases.

In principle, the information given to resettlement agencies should provide a fairly precise estimate of how many cases are expected to arrive.
Indeed, before the start of each fiscal year, the US Department of State announces how many refugees it intends to resettle in that fiscal year, and resettlement agencies are instructed to prepare for a certain fraction of this total number.
In fact, HIAS sets its affiliate capacities to sum up to 110\% of this number of announced refugees, which is intended to give local affiliates a good idea of how many refugees they will receive while affording the resettlement agency some freedom in its allocation decisions.

\subsection{Relying on Capacities}
\label{sec:nfromcaps}
\begin{figure}[tbh]
\centering
\includegraphics[width=\textwidth]{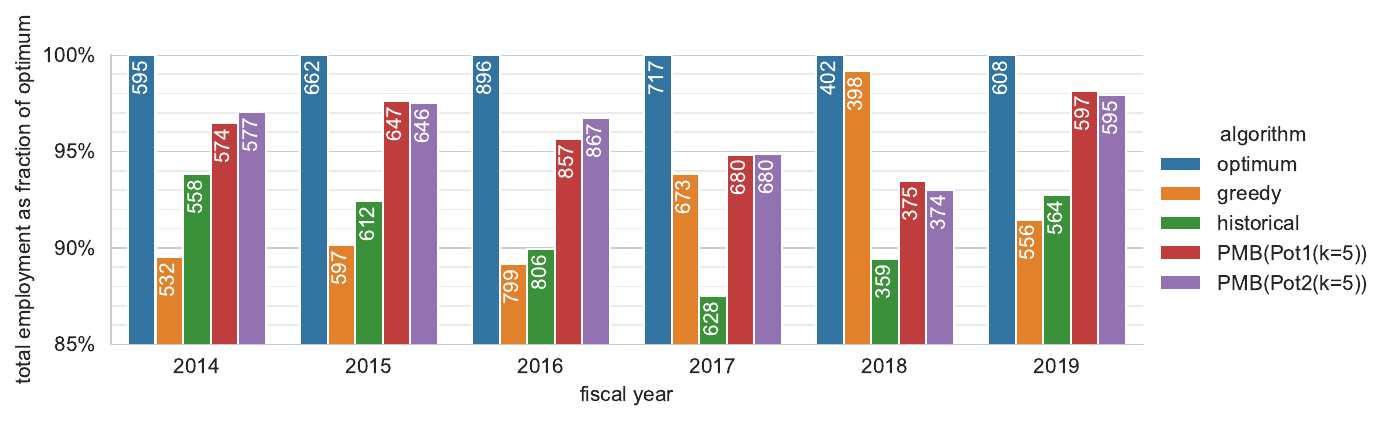}
\caption{Total employment, where cases are not split up and arrive in batches. The potential algorithms no longer have access to the true number of arriving cases but assume that the arriving refugees amount to 91\% of the total capacity. Capacities are the \emph{initial} capacities of the fiscal year (except for historical). For the potential algorithms, the mean employment across 50 random runs is shown.}
\label{fig:trustcaps}
\end{figure}
It is thus natural to run our potential algorithms under the assumption that the number of arriving refugees will be $1/(110\%) \approx 91\%$ of the total announced capacity.\footnote{To convert the number of remaining refugees into a number of cases, we divide by the average case size of recent arrivals (over the years, this average size fluctuates between 2.4 and 2.6). While the number of refugees who have arrived is below 91\% of the total capacity, this gives us a total number of cases $n$ for the algorithms. Once the number of arrivals exceeds 91\% of the total capacity, we make the algorithms assume that the current case is the last to arrive, that is, all subsequently sampled trajectories have length zero.}
The result of this strategy is shown in \cref{fig:trustcaps}.
Since these experiments use the initial, unrevised capacities, the employment scores of the hindsight optimum and the greedy algorithm may differ from those in previous experiments, which used the most revised capacities.\footnote{This means that the comparison to the historical algorithm is not quite on equal terms, since the latter is constrained by a different set of capacities. In all fiscal years except for 2017 and 2018, the final capacities are element-wise larger than the original capacities.}
In all fiscal years other than 2017 and 2018, the imprecise knowledge of future arrivals deteriorates the approximation ratio of the potential algorithms, but the potential algorithms continue to clearly outperform the greedy baseline overall, and they outperform the historical matching in every single year.

\begin{figure}[tbh]
\centering
\includegraphics[width=\textwidth]{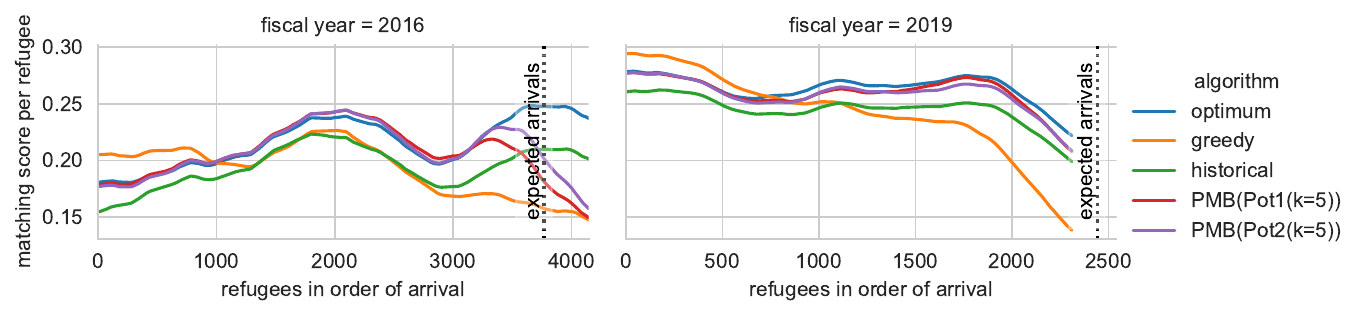}
\includegraphics[width=\textwidth]{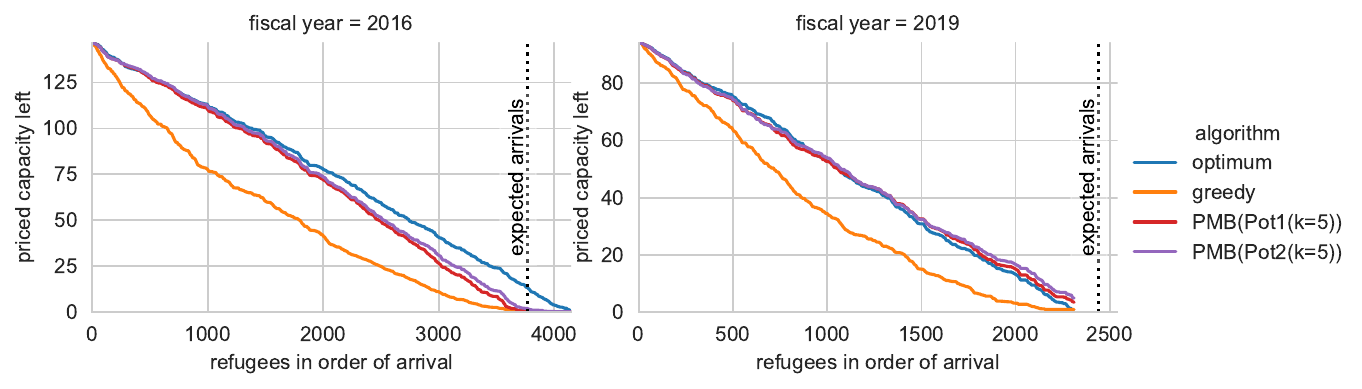}
\caption{Evolution of the per-refugee match score and remaining priced capacity in order of arrival, for fiscal years 2016 and 2019 and one run per algorithm in the experiment of \cref{fig:trustcaps} (whole cases, batches, initial capacities, potential algorithms do not know $n$). Dotted line show how many refugees the potential algorithms expect. Smoothing as in \cref{fig:smoothedunit}. Priced capacity is not shown for historical since it uses different capacities.}
\label{fig:combinedtrusted1619}
\end{figure}
Setting aside the outlier years of 2017 and 2018 for the moment, we investigate the fiscal years 2016 and 2019, in which arrivals were otherwise highest and lowest relative to the announced capacity.
In fiscal year 2016, the total arrivals were particularly large relative to the initial capacity:
the arrival numbers added up to 100\% of the initial capacity rather than 91\%, which means that our potential algorithms expected around 3\,770 refugees to arrive rather than the 4\,150 that ended up arriving.
As a result, the potential algorithms consume the priced capacity at an approximately constant rate, consuming it all around the expected number of expected refugees (\cref{fig:combinedtrusted1619}, bottom left).
Up to this point, the potential algorithms are more generous in consuming capacity than would be ideal given the actual number of arriving cases, which is why the potential algorithms obtain a slightly higher average employment over the first three quarters of arrivals (\cref{fig:combinedtrusted1619}, top left) than the optimal matching in hindsight.
For refugees arriving after the 3\,770 expected refugees, however, the capacity in the best affiliates is used up, which is why the averaged employment sharply drops after this point.\footnote{Note that, due to the triangle smoothing, the drop starts dragging down the curve 500 arrivals before its actual start.}

In 2019, by contrast, fewer refugees arrived than expected, only 86\% of the total capacity.
At the bottom right of \cref{fig:combinedtrusted1619}, it is visible that the potential algorithms consume priced capacity at a slightly lower rate than the optimal algorithm in hindsight, as they aim to use up the capacity around 2\,440 refugees rather than the 2\,310 who ended up arriving.
This effect is reflected in the average employment rates (top right), which lie below that of the optimal algorithm throughout most of the year.\footnote{The drop in employment probabilities at the end of the fiscal year affects all algorithms including the hindsight optimum and must therefore be caused by an anomaly in arrival characteristics.}

The fiscal years of 2017 and 2018 stand out due to the fact that the total number of arriving refugees fell far short of the announced number reflected in the approved capacities: in 2017, arrivals amounted to 65\% of the approved capacities, while they amounted to only 46\% in 2018.
Both of these years fall into the beginning of the Trump administration, which not only sharply reduced the announced intake of resettled refugees, but furthermore abruptly halted the intake of refugees from six predominantly Muslim countries starting from early 2017.
\begin{figure}[tbh]
\centering
\includegraphics[width=\textwidth]{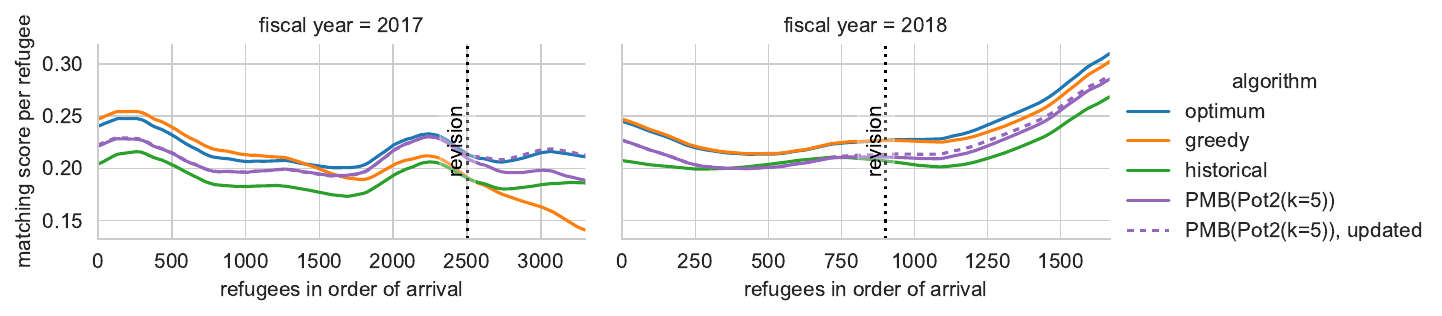}
\includegraphics[width=\textwidth]{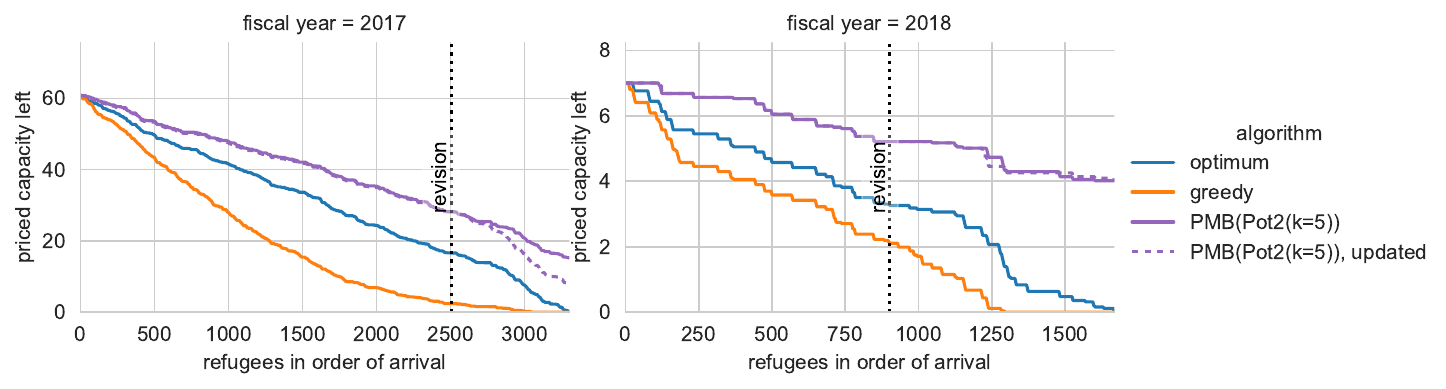}
\caption{Evolution of the per-refugee match score and remaining priced capacity in order of arrival, for fiscal years 2017 and 2018 in the experiment of \cref{fig:trustcaps} (whole cases, batches, initial capacities, potential algorithms do not know $n$).
Dashed line shows evolution if potential algorithm updates its expected arrival number at time of capacity revision (dotted line).}
\label{fig:combinedtrusted1718}
\end{figure}

As the potential algorithm depicted in \cref{fig:combinedtrusted1718} severely overestimates how many cases will arrive, it holds back much more priced capacity than would be optimal (bottom, solid lines).
This causes the potential algorithms to extract less employment throughout the year than the optimal algorithm (top, solid lines).
As observed in \cref{sec:onlinebipartiteevaluation}, the capacities in 2018 are so loose that the greedy algorithm performs close to optimal.

In these two years, the US Department of State eventually reacted by correcting the expected arrivals downward and instructing the resettlement agencies to reduce their capacities.
In fiscal year 2017, this revision came quite late and ended up underestimating the arrivals:
where the arrivals amounted to only 65\% of the initial capacities, they exceeded the revised total capacity at a level of 103\%, rather than amounting to the 91\% that was intended.
Even if imperfect, this signal that arrivals are much lower than originally announced is still useful to the potential algorithms.
Indeed, in \cref{fig:combinedtrusted1718}, the dashed curve corresponds to a potential algorithm that still starts out expecting 91\% of the initial capacities to arrive, but expects only 91\% of the revised capacities to arrive from the point on where they were announced (vertical line).
While this information comes late, the algorithm in fiscal year 2017 uses the new information to burn through the remaining priced capacity more aggressively (bottom left), which allows for higher employment among refugees arriving after the revision of arrival numbers (top left).
As a result, the employment reaches 97\% of the optimum in hindsight, exceeding the value of 95\% without the updated information that we showed in \cref{fig:trustcaps}.

By contrast, the revision in fiscal year 2018 did not yield much useful information; whereas the arrivals amounted to 46\% of the initial capacities, they still amounted to 48\% of the revised capacities.
This seems to indicate that, even after half of the fiscal year's refugees had already been allocated, the administration overestimated the number of arriving refugees by a factor of two.
Because the revision barely changed the number of expected arrivals, giving the potential algorithm access to this revised information does not have much effect (\cref{fig:combinedtrusted1718}, right).

While we have considered the informational value of revisions above, our experiments have not considered that these revisions actually reduced the allowable capacities.
Although we include a variant of the experiment in \cref{app:changingquotas}, it is difficult to meaningfully compare the employment achieved by different algorithms if the parameters of the matching problem are changed so drastically during the matching period.
One particular challenge is that, while the amount of reduction was extraneously decided, HIAS was involved in deciding which capacities to decrease, which was done in a way that depended on previous allocation decisions.\footnote{While the sum of capacities did not change much in fiscal year 2018, the capacities of some affiliates were substantially decreased and those of others were substantially increased.}
Since we only know the revised capacities that were agreed upon, not the counterfactual revision of capacities that would be made, the greedy algorithm and the potential algorithms might have already exceeded a reduced capacity before it was announced.
This means that the experiment rewards algorithms for greedily using up the capacity in the best affiliates before the revision, which we do not expect to be a good policy in practice.
More generally, a substantial change in capacities is an exceptional situation, outside of our model, and cannot be addressed by our algorithm alone without manual intervention.

{
\begin{figure}[tbh]
\centering
\includegraphics[width=\textwidth]{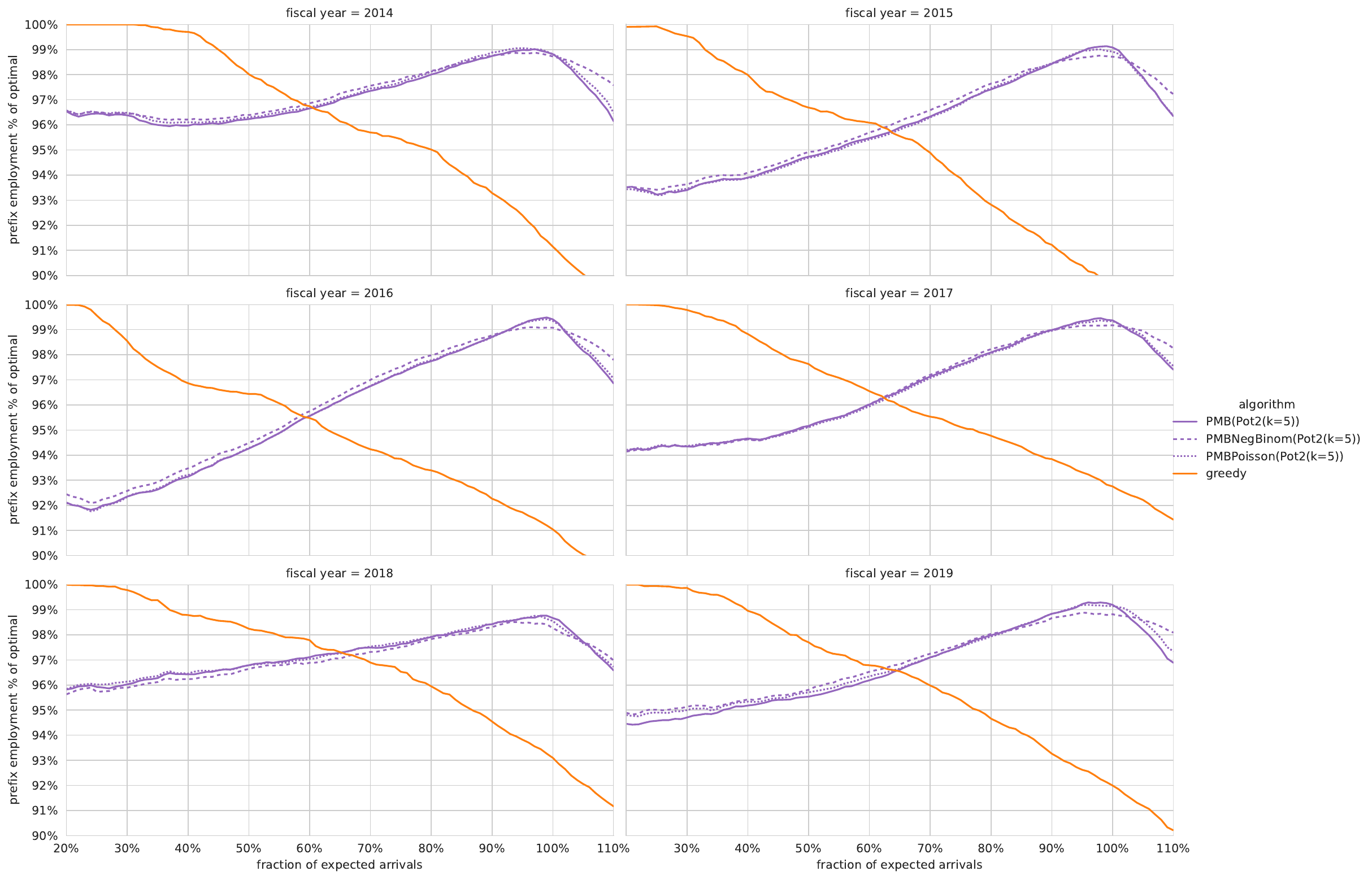}
\caption{Employment achieved by different algorithms as a function of how many refugees arrive. Refugee arrivals are bootstrapped over each fiscal year's historical arrivals, and the number of arriving refugees are given as a fraction of the historical arrivals. Capacities are 110\% of historical matching. Employment is measured as a ratio of the optimal hindsight employment for the same set of arriving refugees. Curves are averaged over 10 arrival sequences.}
\label{fig:bootstrap}
\end{figure}

\subsection{Arrival Misestimation on Bootstrapped Data and Incorporating Uncertainty}
\label{sec:bootstrap}
To obtain more systematic insights into the robustness of potential algorithms to misestimated arrival numbers, we study bootstrapped case arrivals, which allows us to simulate varying numbers of arrivals.
The results of this experiment are displayed in \cref{fig:bootstrap}.
As a baseline, consider the greedy algorithm, which obtains optimal employment when the number of arrivals is much lower than the total capacity (say, 25\% of the expected arrivals, which is $\frac{25\%}{110\%} \approx 23\%$ of the capacity), but becomes more and more suboptimal the more refugees arrive.

By contrast, the potential algorithms perform best (around 99\% of the optimal employment) when the number of arriving refugees matches what the algorithm expects.
On average, this number is around half of a percentage point higher than in the corresponding non-bootstrapped experiments (\cref{fig:full}).
Such an increase is to be expected as the bootstrapping setup ensures that the algorithm draws trajectories from the same distribution from which the arrivals are generated.
In particular, the real arrival sequence used for \cref{fig:full} might contain a drift in refugee characteristics or a seasonality not captured by our algorithm, and the lack of these features in the bootstrapped experiment allows for slightly higher employment.
It is just as noticeable, however, that this increase is \emph{only} half a percentage point, revealing that a drift of arrival characteristics and seasonality does not account for most of the remaining optimality gap of our algorithm.

The further the actual arrival number deviates from this expectation, the further the relative employment performance of the potential algorithm decreases.
Noticeably, the performance more quickly deteriorates when the arrival numbers exceed the expectation, versus falling short.
This sharp decline makes sense for two reasons.
First, the algorithms aim to exploit all useful capacity exactly at the expected number of refugee arrivals; thus, only a subset of the affiliates remain available for subsequent arrivals.
Second, once the number of arrivals exceeds the expectation, the trajectories in the potential algorithms add no cases beyond those that have already arrived, which means that the algorithm serves subsequent arrivals greedily.
In the six fiscal years we observe, arrivals below the expectation seem like a more urgent problem than arrivals above the expectation, but over-arrivals might well become a problem under different political circumstances or when applying potential algorithms to other matching settings.

A natural way to make the potential algorithms more robust to inaccurate arrival estimates is to treat arrival estimates not as exact predictions but as subject to some uncertainty.
Concretely, we adapt the potential algorithms by sampling trajectories of different lengths, each drawn from a ``prior'' distribution whose mean is the arrival estimate, conditioning this distribution such that trajectory lengths are never less than the number of refugees who have already been allocated.
Conceivably, these adapted trajectories could generate potentials that are robust across a wider range of arrival numbers, and the adapted algorithm could therefore lead to higher employment when the official arrival numbers are inaccurate.
The most obvious distribution is perhaps a Poisson distribution.
As shown by the dotted line in \cref{fig:bootstrap}, using Poisson trajectories hardly changes the employment outcomes for any of the experiments relative to the baseline of fixed trajectory sizes.
This is most likely due to the low variance of the Poisson distribution.
For a quite typical mean of 3\,000 arriving refugees, 95\% of the probability mass lies within a distance of only 3.6\% of the mean.
For this reason, we also try a distribution with overdispersion, specifically a negative binomial distribution parameterized to have its mean equal to the expected arrivals and its standard deviation equal to 10\% of the expected arrivals.
For example, if again 3\,000 arrivals are expected, 95\% of the probability mass deviates up to 20\% from the mean.
As the figure shows, negative-binomial trajectories lead to decent improvements in employment when more refugees arrive than expected.
When fewer refugees arrive than expected, using random trajectory lengths helps more often than not, though with different degrees of success.
Overall, negative-binomial arrivals seem to make the potential algorithms marginally more robust to misestimated arrival numbers, though not by enough to make misestimation less of an overall concern.
Additionally, this additional robustness comes at a nonnegligible cost when arrival estimates are accurate.
}

\subsection{Better Knowledge of Future Arrivals}
\label{sec:bettern}
In \cref{sec:nfromcaps}, we demonstrated that, even without outside supervision, our potential algorithms lead to substantial employment increases over the baselines, unless the announced capacities miss the eventual arrival numbers by an extreme margin.
Even in these typical years, however, more accurate arrival predictions could increase the total employment on the order of percentage points of the hindsight optimum.
Obviously, more accurate information about arrivals would be even more useful in years like 2017 and 2018, in which the official information is unreliable.

\begin{figure}[tb]
\centering
\includegraphics[width=\textwidth]{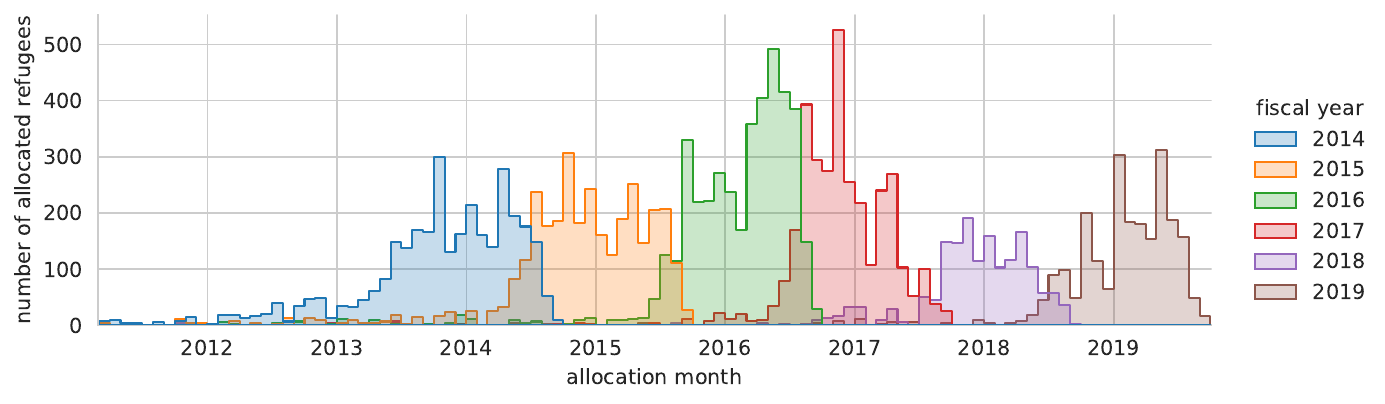}
\caption{Monthly number of allocated refugees, disaggregated by fiscal year of arrival.}
\label{fig:arrivals}
\end{figure}
One approach would be to use time-series prediction to estimate the number of arrivals.
For instance, when the US Department of State revised the capacities for the fiscal year 2018 in January 2018 (several months into the fiscal year), the announcement that 2.5 times more refugees were still to come than had already arrived might have raised some doubts.
However, the graph of monthly arrivals in \cref{fig:arrivals} shows that late increases in arrival rates may actually happen as they did in fiscal year 2016.\footnote{In fiscal year 2016, the number of arrivals after January 2016 was 1.6 times larger than the number that had arrived so far. In the fiscal year of 2015, the number of refugees arriving after January 2015 was only 75\% of that arriving before.}

A fundamental challenge that any data-driven approach faces is that there is very little data to learn from.
Indeed, while HIAS has data on hundreds of thousand of refugees, they only have data on 15 fiscal years, which is, moreover, incomplete and smaller-scale in earlier years.
Thus, there is a limited foundation to learn about how arrival patterns change between years.
This task becomes especially difficult given that arrival numbers are heavily influenced by external events such as elections, the emergence of humanitarian disasters, and changes in immigration policy, which cannot be deduced from past arrival patterns. 
Thus, while a time-series prediction approach might lead to marginal improvements over na\"ively expecting 91\% of the capacity to arrive, past arrival numbers are unlikely to give enough information to accurately predict future arrival numbers.

Fortunately, resettlement agencies such as HIAS already possess much richer information and insights into the dynamics of refugee arrivals than a pure data approach would consider.
In fiscal year 2017, for example, HIAS foresaw a worsening climate for refugee resettlement immediately after the November 2016 election\footnote{\url{https://www.hias.org/news/press-releases/hias-calls-president-elect-trump-respect-longstanding-refugee-policy}} and was aware of concrete plans to drastically reduce refugee intake in January 2017,\footnote{\url{https://www.hias.org/news/press-releases/trumps-planned-action-refugees-betrayal-american-values}} both before these changes were reflected in arrival numbers and before the capacities were officially updated in March 2017.
Similarly, HIAS continuously monitors domestic politics and international crises for their potential impact on resettlement, and moreover it has some limited insight into the resettlement pipeline, which allows it to prepare for changes in arrivals.
We therefore believe that, rather than building a sophisticated tool for predicting arrivals in a fully autonomous manner, it is preferable to allow HIAS staff to override our prediction with more advanced information.

\section{Implementation in \annie{} \textsc{Moore}}
\label{sec:implementation}
To enable HIAS to benefit from dynamic allocation via potentials, we have integrated new features into its matching software \annie{} \textsc{Moore}.
A crucial design requirement is that HIAS staff must be able to override the allocation recommendations of \annie{} when they are aware of requirements outside of our model.  %
From an interface-design perspective, the challenge is to visualize the effect of such overrides on total employment, enabling HIAS staff to make informed trade-offs.
In the original, static model, this was easy enough:
as the quality of a matching was just the total employment of the current batch, the interface labeled each case--locality match with its associated employment score, and staff could drag the case to other localities to see the respective employment scores.
In a dynamic setting, however, presenting only the employment scores may unintentionally encourage HIAS staff to greedily use capacity in their overrides, at the expense of future arrivals.

\begin{figure}[tb]
\centering
\includegraphics[width=1\textwidth]{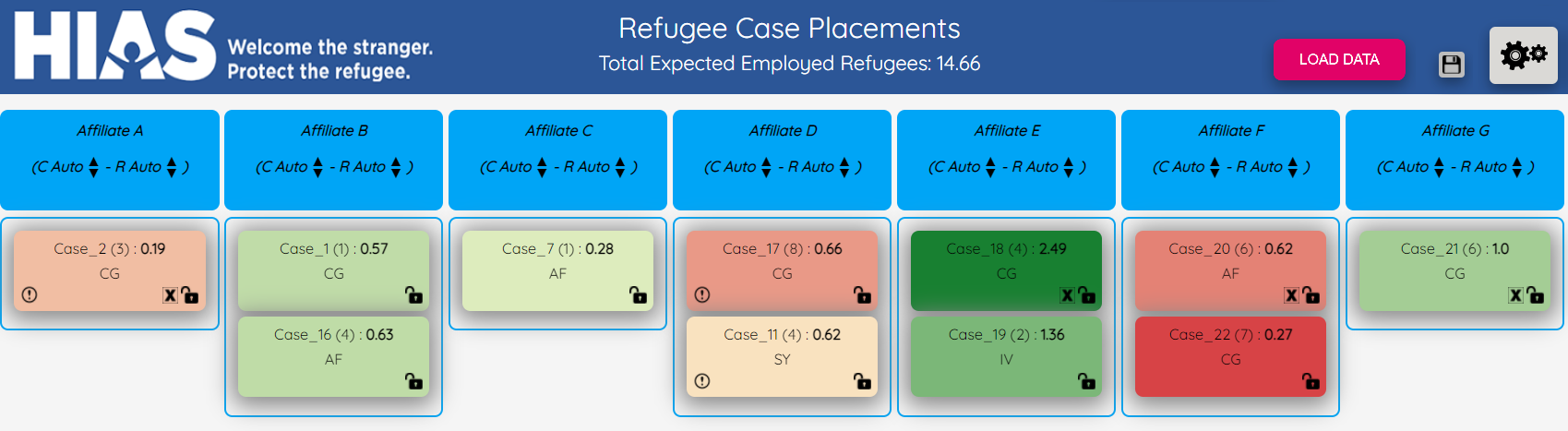}
\caption{Updated \annie\ Interface. Family tiles now show both original numerical employment scores of families in affiliates, as well as the \emph{adjusted} employment score by its shading. Green indicates positive adjusted scores, red negative scores, and darker colors represent greater magnitudes.}
\label{fig:Implementation_new_interface}
\end{figure}

As we illustrate in \cref{fig:Implementation_new_interface}, the new interface of \annie{} augments the original interface with information about affiliate potentials, thereby taking future arrivals into account.
Specifically, the background color of the tile for case $i$ encodes the \emph{adjusted} employment score, that is, the original employment score $u_{i, \ell}$ less the potential $s_i \, p_\ell$ of the capacity consumed in affiliate $\ell$.\footnote{The employment scores of cases in affiliates are prominently retained in a text label.}
The fact that the algorithm \textsf{PMB} always maximizes the sum of adjusted employment scores in its allocation of the current batch means that the algorithm is explainable in terms of the information presented to the user. 
In the interface, the \emph{green} color spectrum indicates positive adjusted employment scores (meaning that the employment score of the case outweighs the loss in future employment), while the \emph{red} color spectrum highlights negative adjusted scores (where a placement reduces future employment by more than its employment score).
Darker colors signify greater magnitudes.

\begin{figure}[tb]
\centering
\includegraphics[width=1\textwidth]{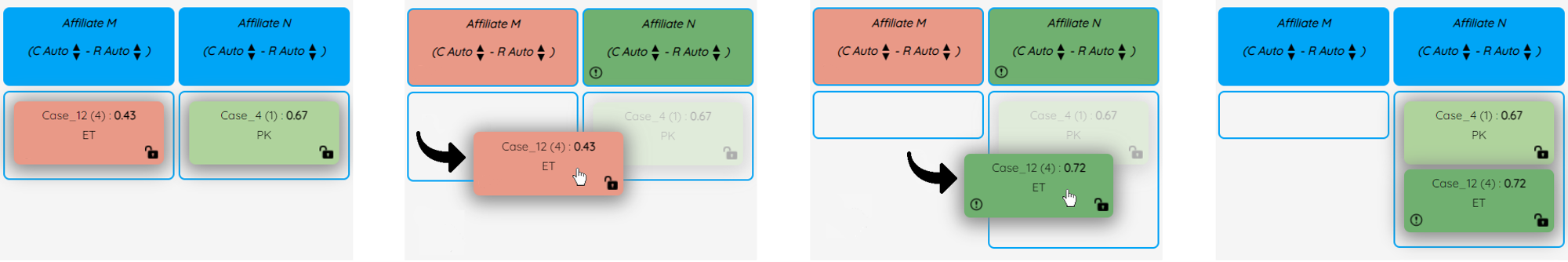}
\caption{Moving a family tile. Other case tiles fade, and affiliate tiles are colored as per their adjusted employment scores in shades of green (positive) or red (negative). Exclamation marks indicate incompatibilities.}
\label{fig:Implementation_Upon_dragging}
\end{figure}

In overriding the allocation recommended by \annie{}, HIAS staff should be able to quickly find alternative placements for a case that do not reduce immediate and future employment by more than necessary.
To support this workflow, our interface shows the adjusted employment scores of a case across all affiliates at a glance:
as shown in \cref{fig:Implementation_Upon_dragging}, upon dragging a particular case tile from its current placement, all other case tiles temporarily fade in appearance, and the shading of every affiliate tile temporarily assumes the adjusted employment score relative to the selected case.
By hovering a selected case tile over a new affiliate, the original (numeric) employment score and the adjusted match score (background color of the case tile) dynamically update.
Moreover, incompatibilities with affiliates due to nationality, language, family size, and single parent households can be seen via an exclamation mark in the lower left corner of the affiliate tile.
After dropping the case tile in a new affiliate, the background color for each affiliate returns to its original blue shade, and all affiliate-tile exclamation marks disappear.

On a separate screen (not shown), \annie{} enables the entry of a prediction for total refugee arrivals, as mentioned in \cref{sec:bettern}.
This estimate can be critical to inform the process of estimating proper shadow prices, as at times HIAS is in a better position to give more accurate case arrival predictions than officially announced capacities.


\section{Conclusion}
\label{sec:conclusion}
We have developed and implemented algorithms for dynamically allocating refugees in a way that promotes refugees' prospects of finding employment.
These algorithms outperform the baselines, even when taking into account how refugee placement in practice deviates from a classic matching setting.

While we have tested the algorithms as an autonomous system, the success of \annie{} in increasing employment outcomes in practice will depend on how it performs in interaction with HIAS resettlement staff.
In \cref{sec:bettern}, we already saw that the allocation decisions of \annie{} can greatly profit from human decision makers providing better estimates of future arrivals.
Human input is equally crucial in dealing with uncertainty in several other places:
for example, HIAS staff might intervene by correcting the arrival year of a case should the Department of State's estimate seem off, or they might increase some affiliate capacities late in the year if they anticipate that these capacities can be increased.
By allowing all parameters of the matching problem to be changed, \annie{} allows HIAS resettlement staff to improve the matching using all available information.

Ideally, the human-in-the-loop system consisting of the matching algorithm and HIAS staff can combine the strengths of both of its parts:
On the one hand, the algorithms in \annie{} capitalize on subtle patterns in employment data and manage capacity more effectively over the course of the fiscal year.
On the other hand, the expert knowledge of HIAS staff enables the system to handle the uncertainty that is inherent in a matching problem involving the actions of multiple government agencies, dozens of affiliates, and thousands of refugees.
In light of the administration's recent increase of the total resettlement capacity from 15\,000 to 125\,000,\footnote{\url{https://www.hias.org/news/press-releases/refugee-cap-fy2022-set-125000}} we foresee both parts playing a crucial role:
the increasing scale of the problem will make data-based algorithms more effective, and human guidance will be necessary to navigate the evolving environment of a rapidly growing operation.

\begin{acks}
We are deeply grateful to HIAS for providing data and for sharing insights into the practical challenges of refugee resettlement. We thank Siddhartha Banerjee, Avrim Blum, Bailey Flanigan, and David Wajc for helpful discussions. This work was supported by Economic and Social Research Council grant ES/R007470/1; by National Science Foundation grants CCF-1733556, CCF-2007080, CMMI-1825348, and IIS-2024287; and by Office of Naval Research grant N00014-20-1-2488.
\end{acks}

\bibliographystyle{ACM-Reference-Format} %
\bibliography{bibliography} %

\appendix

\newpage
\section*{\LARGE Appendix}

\section{Data Preprocessing and Employment Prediction}
\label{app:datacleaning}
We use arrival data directly collected by HIAS, recent as of Summer 2020.
We drop entries in the database missing crucial information: agent and case identifiers, the case's pool (free or with US ties), and the dates of birth, of allocation, and of arrival.
We furthermore remove all cases that were allocated before March 2011 since our employment prediction uses local unemployment data, which is incomplete before then.
This last step removes 3.5\% of refugees from fiscal year 2014, and less than 1\% of refugees from the subsequent fiscal years. 
Due to this removal, we do not include fiscal years before 2014 in our analysis, for which a higher percentage of arrivals would have been removed.

To predict employment scores, we train two separate LASSO models following the methodology of Ahani et al.~\citep{ahani2020placement}, one model for free cases and one for tied cases.
Training separate models is helpful since, as Ahani et al.\ note, tied cases find employment in different patterns due to existing support networks.
The regression is trained on cases allocated in fiscal years 2011 to 2019, where the hyper-parameters are determined via cross validation before retraining on the whole time range.
Note that training on years that we evaluate on would be problematic for evaluating how well the employment scores match the ground-truth employment.
However, since our evaluations measure how well the matching algorithms do relative to the given employment scores, training on a wide range of years ensures that the employment scores are as accurate as possible.

\section{Alternative Potential Approach}
\label{app:potential2}
In \cref{sec:potentialapproach}, we presented a procedure for calculating potentials, termed \textsf{Pot1}, which was based on the shadow prices of a matching LP.
We also outlined a second potential procedure, \textsf{Pot2}, in the body, which we formally define and justify here:
\begin{algorithm}[htbp]
\Parameter{$k \in \mathbb{N}_{\geq 1}$, the number of trajectories per potential computation}
\KwIn{remaining capacities $\bm{c}$, the index $t'+1, \dots, t$ of the cases in the current batch (see \cref{alg:batching}), characteristics of cases arriving in the 6 past months}
\KwOut{a set of potentials $p_\ell$ for all affiliates $\ell$}

\For{$j = 1, \dots, k$}{
for each $i = t+1, \dots, n$, set $s_{i}$ and $\{u_{i, \ell}\}_{\ell}$ to the size and employment scores of a random, recently arrived case\;
solve the following bipartite-matching LP:
\begin{empheq}[box=\fbox]{align*}
    \text{maximize}~& \sum_{i=t'+1}^n \sum_{\ell \in L} u_{i, \ell} \, x_{i, \ell} \\
    \text{subject to}~& \sum_{\ell \in L} x_{i, \ell} = 1 & \forall i = (t'\!+\!1), \dots, n \\
    & \sum_{i = t'+1}^n s_i \, x_{i, \ell} \leq c_{\ell} & \forall \ell \in L & \quad (*) \hspace{3mm} \label{eq:algcapacity} \\
    & x_{i, \ell} \geq 0 & \forall i = (t'\!+\!1), \dots, n, \forall \ell \in L.
\end{empheq} \label{lin:potential2lp}\\
for each $\ell$, set $p_{\ell}^j$ to be the the minimal shadow price of the constraint $(*)$\;
}
set $p_\ell \leftarrow (\sum_{j=1}^k p_{\ell}^j) / k$ for all $\ell$\;
\Return $\{p_\ell\}_{\ell \in L}$\;
\caption{\textsf{Pot2}}
\label{alg:potential2}
\end{algorithm}
the procedure is given in \cref{alg:potential2}, and it can be immediately plugged into \textsf{PMB}. Note that the dual of the linear program in \cref{lin:potential2lp} looks as follows:
\begin{align*}
    \text{minimize}~& \sum_{i \in t'+1}^n y_i + \sum_{\ell \in L} p_\ell \\
    \text{subject to}~& y_i \geq u_{i, \ell} - s_i \, p_\ell & \forall i = (t'\!+\!1), \dots, n, \forall \ell \in L \\
    & p_\ell \geq 0 & \forall \ell \in L.
\end{align*}

Suppose that $s_i = 1$ for all $i \in N$, as we do in \cref{sec:unit}.
Then, for any (non-fractional) optimal matching $x_{i, \ell}$ in the primal, and any set of optimal dual variables $p_\ell$, it is well known (and follows from complementary slackness) that the two form a \emph{Walrasian equilibrium}.
That is, if we consider $u_{i, \ell}$ as case~$i$'s utility for being matched to affiliate $\ell$, and if we imagine charging case~$i$ a price of $p_\ell$ for being matched to affiliate $\ell$, the optimal matching matches each case $i$ to an affiliate $\ell$ maximizing the profit $u_{i, \ell} - p_\ell$.
Thus, if all trajectories perfectly predicted future arrivals, and if ties were broken in a specific way in each step, equipping \textsf{PM} with the potentials of \textsf{Pot2} would lead to the optimal-employment matching.

Note that this justification also extends to batching (still assuming $s_i = 1$), since the Walrasian equilibrium shows that it is possible to maximize $u_{i, \ell} - p_i$ for all cases at once.
Thus, if the trajectory perfectly anticipates future arrivals, each optimal hindsight matching maximizes the matching ILP allocating each batch in \textsf{PMB}.

When the shadow prices are chosen as the element-wise minimal set of shadow prices, work by Hsu, Morgenstern, Rogers, Roth, and Vohra~\citep{HMR+16} gives arguments that \textsf{PM} with \textsf{Pot2} should be somewhat robust to tie breaking, assuming that employment scores satisfy a genericity condition.
In addition, they show that, for large matching problems, this matching algorithm should also be robust if trajectories are drawn from the same i.i.d.\ distribution as the real set of future arrivals.\footnote{This mapping is not perfect since Hsu et al.\ show that capacities are not exceeded by much, rather than considering cascading effects of refugees switching affiliates once their most preferred affiliate is at capacity. This problem could be remedied by slightly reducing capacities in the computation of shadow prices.}

Like the justification of the \textsf{Pot2} potentials, this justification does not cleanly extend to cases of non-unit size.
Furthermore, there is no immediate theoretical justification for averaging these prices over multiple trajectories, which has to be evaluated empirically.

\section{Pseudocode for algorithm \textsf{PMB}}
\label{app:alg:batching}
\begin{algorithm}[H]
\Parameter{a subroutine $\textsf{Potential}$ to determine affiliate potentials}
initialize the capacities $c_\ell$ for each affiliate $\ell$\;
$t_\mathit{last} \leftarrow 0$\tcp*{index of last case in previous batch}
\While{$t_\mathit{last} < n$}{
observe the size $b$ of the current batch\;
$t \leftarrow t_\mathit{last} + b$\tcp*{index of last case in current batch}
observe the case size $s_i$ and the employment scores $\{u_{i, \ell}\}_{\ell}$ for all $i=(t_\mathit{last}\!+\!1), \dots, t$\;
call $\textsf{Potential}()$ to define a potential $p_{\ell}$ for each affiliate $\ell$\;
let $\{\hat{x}_{i, \ell}\}$ be an optimal solution to the following bipartite-matching ILP with knapsack constraints:
\begin{empheq}[box=\fbox]{align*}
    \text{maximize}~& \sum_{i=t_\mathit{last}+1}^{t} \sum_{\ell \in L} (u_{i, \ell} - s_i \cdot p_\ell) \, x_{i, \ell} & \\
    \text{subject to}~& \sum_{\ell \in L} x_{i, \ell} = 1 & \forall i = (t_\mathit{last} + 1), \dots, t \\
    & \sum_{i = t_\mathit{last} + 1}^{t} s_i \, x_{i, \ell} \leq c_{\ell} & \forall \ell \in L&\hspace{3mm} \\
    & x_{i, \ell} \in \{0, 1\} & \forall i = (t_\mathit{last} + 1), \dots, t, \forall \ell \in L.
\end{empheq}\\
\For{$i=t_\mathit{last}+1, \dots, t$}{
allocate case $i$ to unique affiliate $\ell$ where $\hat{x}_{i, \ell} = 1$\;
$c_\ell \leftarrow c_\ell - s_i$\;
}
$t_\mathit{last} \leftarrow t$\;
}
\caption{\textsf{PMB}(\textsf{Potential})}
\label{alg:batching}
\end{algorithm}

\section{Additional Experiments}
\subsection{Employment Statistics for Non-Unit Cases without Batching}
\label{app:figs:nobatching}

\begin{figure}[H]
\centering
\includegraphics[width=\textwidth]{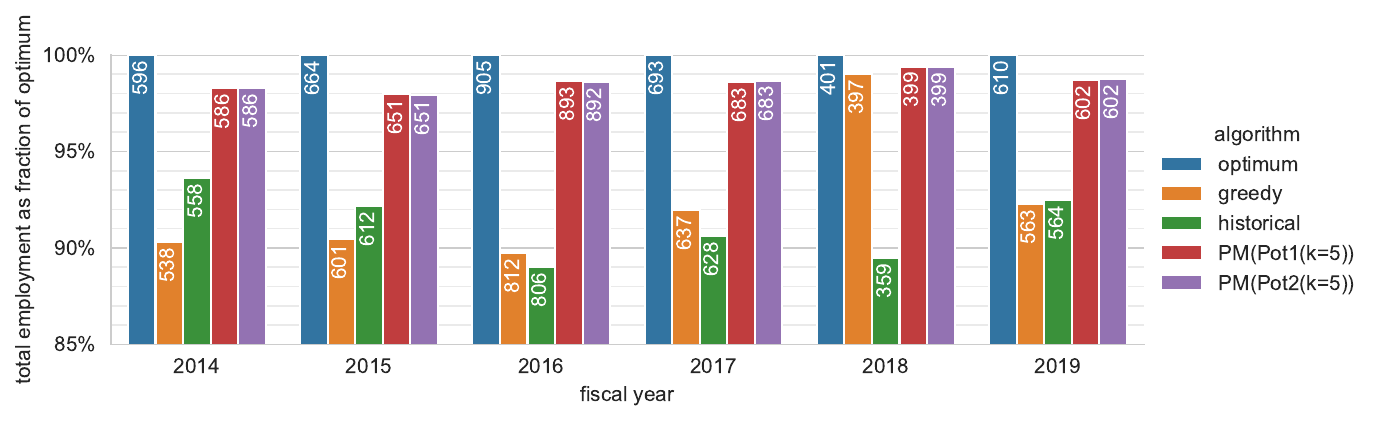}
\caption{Total employment obtained by different algorithms, with whole cases arriving rather than being split up. Capacities are the final capacities of the fiscal year. For the potential algorithms, employment is averaged over 10 random runs. The numbers in the bars denote the absolute total employment; the bar height indicates the proportion of the optimum total employment in hindsight.}
\label{fig:nobatching}
\end{figure}
\begin{figure}[H]
\centering
\includegraphics[width=\textwidth]{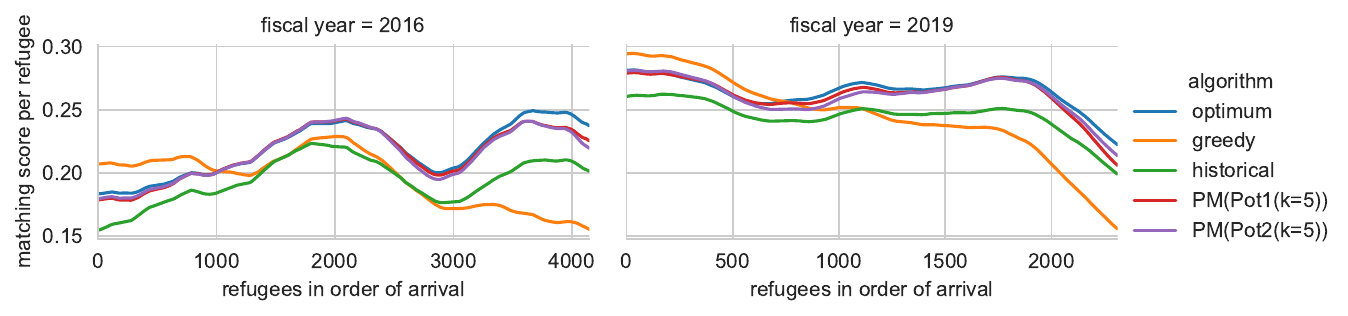}
\caption{Evolution of the per-refugee match score in order of arrival, for fiscal years 2016 and 2019 in the experiment of \cref{fig:nobatching} (whole cases, final capacities). Match scores are smoothed using triangle smoothing with width 500.}
\label{fig:smoothednobatching}
\end{figure}
\begin{figure}[H]
\centering
\includegraphics[width=\textwidth]{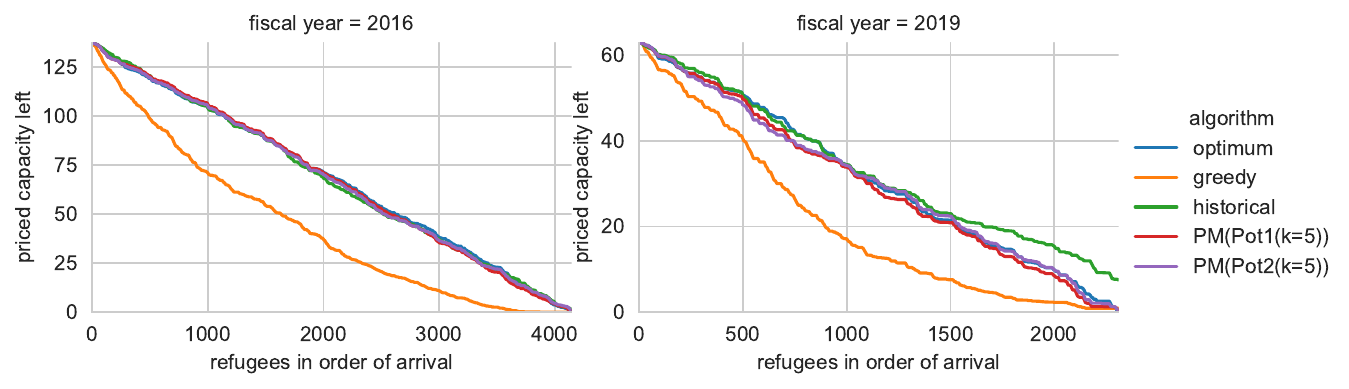}
\caption{Remaining priced capacity at the time of arrival of different refugees, for fiscal years 2016 and 2019 and one random run per algorithm in the experiment of \cref{fig:nobatching} (whole cases, final capacities).}
\label{fig:pricedcapsnobatching}
\end{figure}

\subsection{Timing}
\label{app:timing}
In the experiment of \cref{sec:fullempirics}, we measure the running time for a call to \textsf{PMB}, which mainly consists of the time of sampling the $k$ random trajectories, of computing the appropriate shadow prices for each trajectory, and of solving the matching ILP.
This time will generally increase the more arrivals are expected in the remainder of the fiscal year since this determines the length of the trajectory and the size of the LP whose duals must be computed.
Additionally, batches consisting of more cases should also increase running time since the matching ILP is larger for such batches.

Therefore, we benchmark the algorithms on fiscal year 2016, which saw the largest number of arriving cases (1\,628).
Within this fiscal year, we aim to pick a batch that is both early and large, which is a bit complicated by the fact that the earliest batches of the year are small.
We choose a batch of 31 cases, which comes early enough that 1\,486 of the 1\,628 cases of the fiscal year come after it.
This batch is the 17th largest out of the 168 batches of the year, and all preceding batches are less than half as many cases.

For this batch, we measure the running time for \textsf{PMB} with potentials \textsf{Pot1} and \textsf{Pot2}, each with $k \in \{1, 5, 9\}$.
Each reported running time is the average of 50 iterations and is measured on a 2017 MacBook Pro with a 3.1 GHz Dual-Core Intel Core i5 processor using Gurobi for solving LPs and ILPs:
\begin{center}
    \begin{tabular}{lc}
    \toprule
         Matching algorithm & Running time \\
             \midrule
         \textsf{PMB}(\textsf{Pot1}($k=1$)) & 0.5\,s \\
         \textsf{PMB}(\textsf{Pot1}($k=5$)) & 2.4\,s \\
         \textsf{PMB}(\textsf{Pot1}($k=9$)) & 4.0\,s \\
         \textsf{PMB}(\textsf{Pot2}($k=1$)) & 0.5\,s \\
         \textsf{PMB}(\textsf{Pot2}($k=5$)) & 2.6\,s \\
         \textsf{PMB}(\textsf{Pot2}($k=9$)) & 4.6\,s \\
    \bottomrule
    \end{tabular}
\end{center}
\vspace{.2cm}
Given that a resettlement agency needs to execute the matching algorithm at most once per day, these times are negligible.
For low $k$ and an optimized implementation, it might even be possible to support near real-time experimentation; for example, an \annie{} user could drag a slider to indicate the expected number of arriving refugees, and could see how this would impact the recommended allocation of the current batch.

{
As we mention in the conclusion of the paper, we anticipate that refugee arrival rates will substantially increase in the future.
With this scenario in mind, we investigate how the runtime of potential algorithms scales for much larger arrival numbers than those currently encountered.
Below, we measure the runtime of our algorithm \textsf{PMB}(\textsf{Pot2}($k=5$)) for bootstrapped arrivals based on fiscal year 2016.
In each experiment, we fix a number of arrivals and measure the running time for allocating a simulated batch containing a $\frac{1}{52}$ fraction of the total arrivals, arriving right at the beginning of the fiscal year when the largest LPs must be solved.
We proportionally scale the affiliate capacities based 110\% of the historically allocated capacity, and average each running time over 3 runs.
\begin{center}
    \begin{tabular}{rr}
    \toprule
         Number of arriving refugees & Running time \\
             \midrule
        10\,000 & 9\,s \\
        20\,000 & 25\,s \\
        40\,000 & 70\,s \\
        80\,000 & 4\,min \\
        125\,000 & 12\,min \\
    \bottomrule
    \end{tabular}
\end{center}
\vspace{.2cm}
These numbers suggest that the runtime of the algorithm does not scale unreasonably fast in the number of arrivals.
It would even be feasible to allocate the planned capacity of 125\,000 refugees for the entire US through our algorithm.
Should runtime considerations become a problem, computation could be easily sped up by computing the duals for the different trajectories in parallel.
}

{
\subsection{Length of sampling window}
\label{app:lookback}
Throughout the paper, our potential mechanisms simulate trajectories by bootstrapping cases that were allocated by the algorithm within the previous six months, a duration to which we will refer as the algorithm's \emph{sampling window}.

The sampling window of six months was \emph{not} selected based on data, which is important to avoid overfitting the data.
Instead, we chose the duration of six months as a reasonable middle ground, which we hoped would be short enough so that the algorithm may react to a sudden shift within the same matching period and simultaneously would be long enough that, at any point, the algorithm would have a sufficiently diverse and representative set of recently allocated cases.

\begin{figure}[tbh]
\centering
\includegraphics[width=\textwidth]{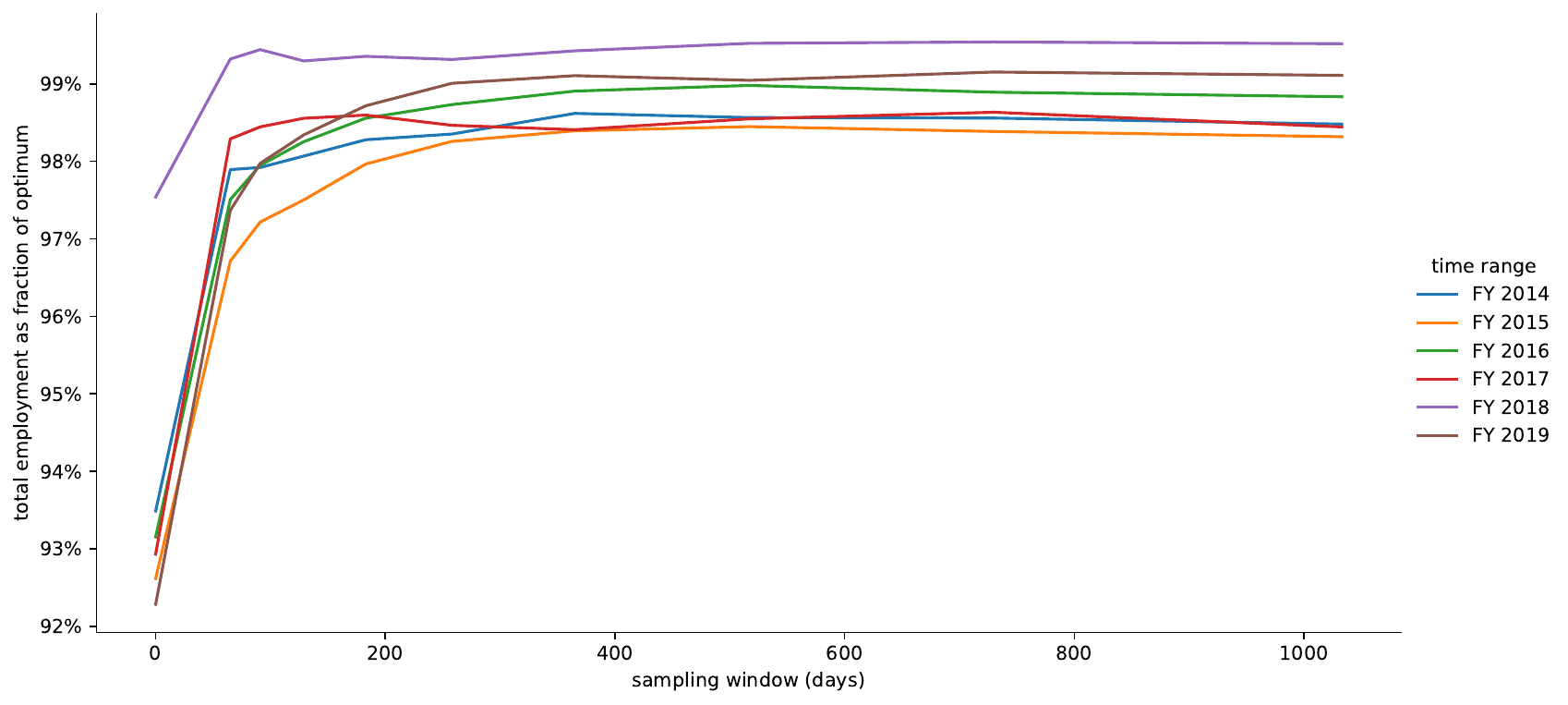}
\caption{Employment of algorithm \textsf{PMB(Pot2(}$k=5$\textsf{))} as a function of the sampling window. Capacities are the final fisyear capacities, and the experiment is comparable to \cref{fig:full}. Each datapoint is averaged over 10 random runs.}
\label{fig:lookback}
\end{figure}
Though our concern for overfitting compels us not to optimize the sampling window for use in our main results, we can explore in \cref{fig:lookback} how different sampling windows would have impacted our algorithm's performance.
As should be expected, a sampling window of length 0 harms employment performance since, in this case, trajectories are bootstrapped form only the cases in the current batch and thus prices are of little value.
Already the second-smallest sampling window we evaluated, 63 days, leads to much better employment.
Indeed, in the fiscal years 2014, 2017, and 2018, this performance is already close to optimal, whereas in the remaining three fiscal years, larger sampling windows further increase the employment performance by a fair amount.
To our surprise, very long sampling windows between 1.5 and 3 years do not seem to lead to lower performance than intermediate sampling windows such as the one we chose.
It seems that, despite dramatic shifts in refugee origin and admission policy, changes in refugee characteristics over time change little in the calculus of how valuable each affiliate is, which is in line with our observations in \cref{app:pricesbyarrivals}.
In light of these results, our own choice of 183 days seems reasonable, but we would cautiously correct this number upward, to perhaps one year.
Even though we have not observed any case where very long sampling windows would be bad for employment, this finding based on only six years of data should not be overinterpreted---it is still plausible that future shocks in refugee characteristics will require a shorter sampling window to allow for faster adaptation.

\subsection{Evolution of potentials by arrival numbers} 
\label{app:pricesbyarrivals}

\begin{figure}
    \centering
    \includegraphics[width=\textwidth]{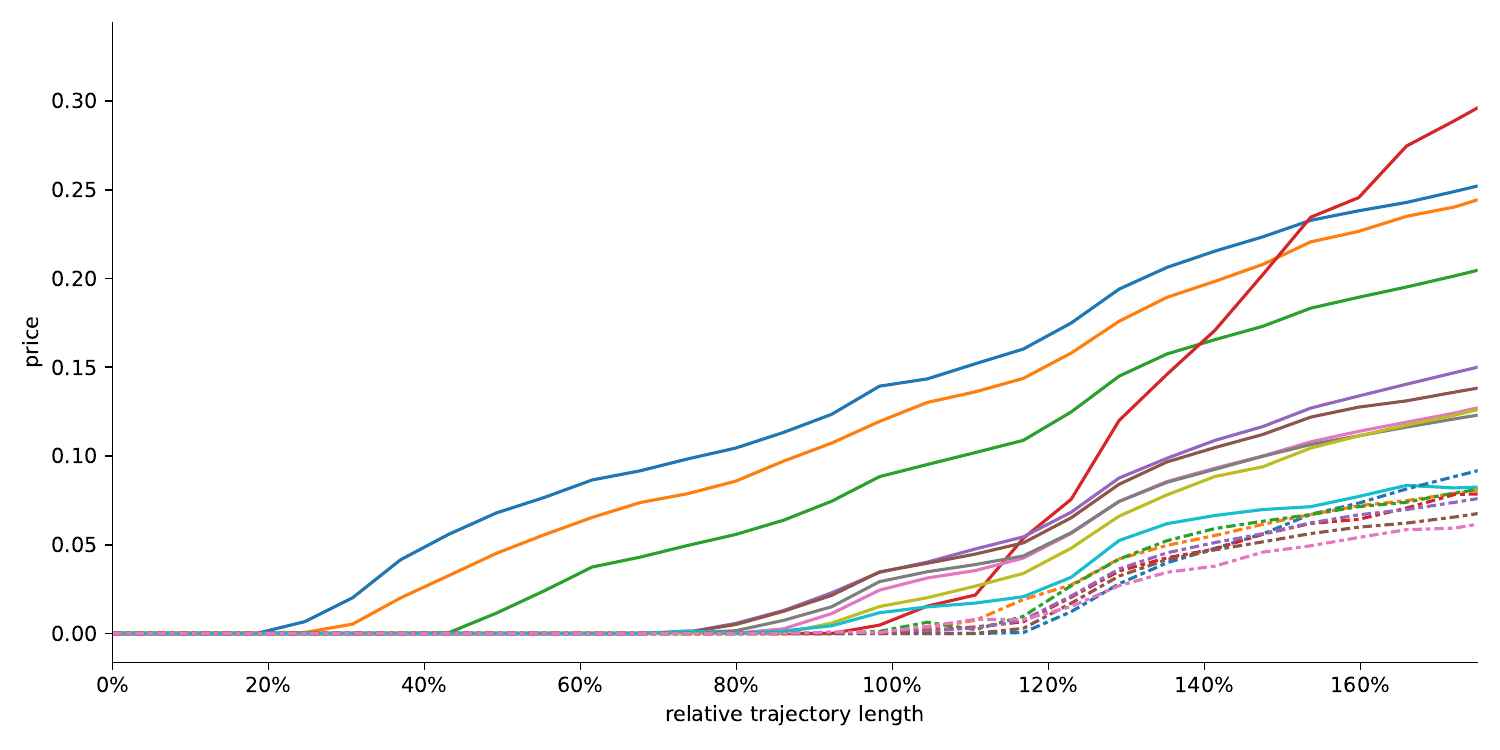}
    \caption{At the beginning of fiscal year 2016, the $\textsf{Pot2}$ potentials of different affiliates depending on the length of simulated trajectory. The trajectory length is measured as a number of cases and given as a percentage of the actual number of case arrivals in fiscal year 2016. Each line corresponds to an affiliate, which is not identified due to privacy considerations. Trajectories are bootstrapped over the fiscal-year 2016 arrivals, and the lines averaged over 50 random trajectories. Five affiliates with capacities below 15 are omitted.}
    \label{fig:potentialsbylength}
\end{figure}

\Cref{fig:potentialsbylength} shows how the number of arriving cases impacts the prices of different affiliates.
The figure shows data for fiscal year 2016 and the potential $\textsf{Pot2}$; the corresponding plots for other fiscal years and for $\textsf{Pot1}$ look quite similar.
For very short trajectories, all potentials are zero, which reflects that there is no scarcity of affiliate capacity and that greedy assignment is optimal.
As the trajectory length increases, the potentials of more and more affiliates become positive.
Generally speaking, affiliate potentials increase with the trajectory length, and they rarely cross, which means that, even across very different arrival numbers, which affiliates' capacities are most valuable remains stable and roughly corresponds to the average employment over all refugees.

One affiliate deviates from this pattern across fiscal years, namely the one represented by the solid red line in \cref{fig:potentialsbylength}. %
This affiliate has a low potential up to the historical trajectory length, but, for large trajectory lengths, its potential increases steeply and becomes the largest overall.
This affiliate stands out in three ways: (1) a large number of refugees (in 2016, amounting to 82\% of the affiliate's capacity) are compatible with only this affiliate, (2) these tied refugees have very high per-person employment prospects (36\%), and a vast majority of the other cases are not compatible with this affiliate. %
Thus, when the trajectory length is short, all tied cases fit in the affiliate and the compatibility constraints mean that there is little competition for the remaining capacity among non-tied cases.
This changes when the trajectory length is long enough that the number of tied refugees exceeds the capacity: given that not all tied refugees can be matched, a high potential ensures that the slots are used for the refugees with largest employment opportunities.
While the above observations are technically interesting and might be useful in other applications of online matching, we stress that they refer to extreme scenarios in which much more refugees arrive than in any of our data, and in which the affiliate's capacity is dramatically at odds with the demand from tied cases.

\subsection{Robustness checks using other arrival sequences}
\label{app:reversedetc}
\Cref{fig:full} shows that the potential algorithms consistently achieve between 98\% and 99\% of the optimal employment in hindsight, a clear step up from the greedy matching and the matching historically chosen by HIAS.
That being said, the evaluation in the main text is limited to six different arrival sequences, one per fiscal year observed.
To demonstrate the robustness of our findings, we will repeat the experiment with two other forms of arrival sequences, derived from but not identical to the historical arrival sequences.

\begin{figure}
    \centering
    \includegraphics[width=\textwidth]{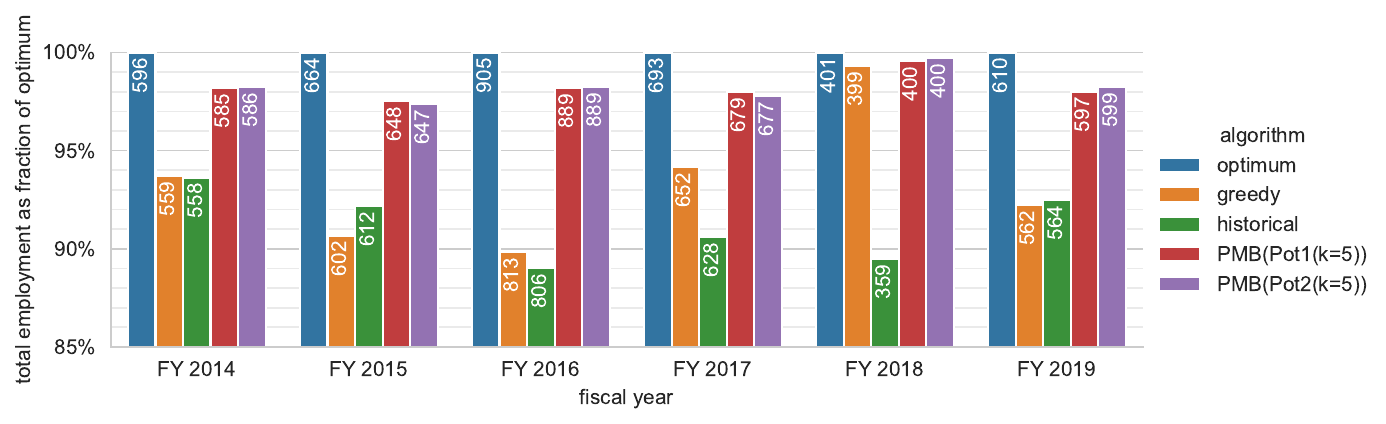}
    \caption{Version of the experiment in \cref{fig:full}, where each fiscal year's arrival sequence is reversed. For the potential algorithms, total employment is averaged over 10 random runs.}
    \label{fig:reversed}
\end{figure}
First, in \cref{fig:reversed}, we reverse the order in which each fiscal year's arrivals are presented to the algorithm.
Note that the optimal and historical matching do not change in this case.
Remarkably, the performance of the greedy algorithm improves overall, by as much as 3.5\% in 2014 and 1.7\% in 2017.
In contrast to greedy, the performance of the potential algorithms seems to change much less.
While there seems to be a trend of the potential algorithms performing slightly better forward than backward, these decreases are clearly below one percentage point.

\begin{figure}
    \centering
    \includegraphics[width=\textwidth]{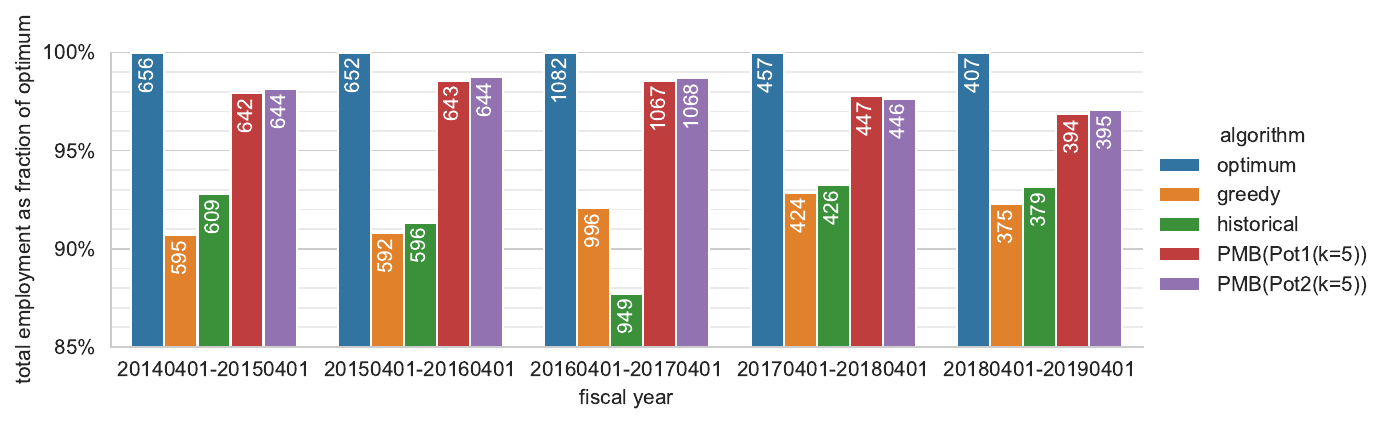}
    \includegraphics[width=\textwidth]{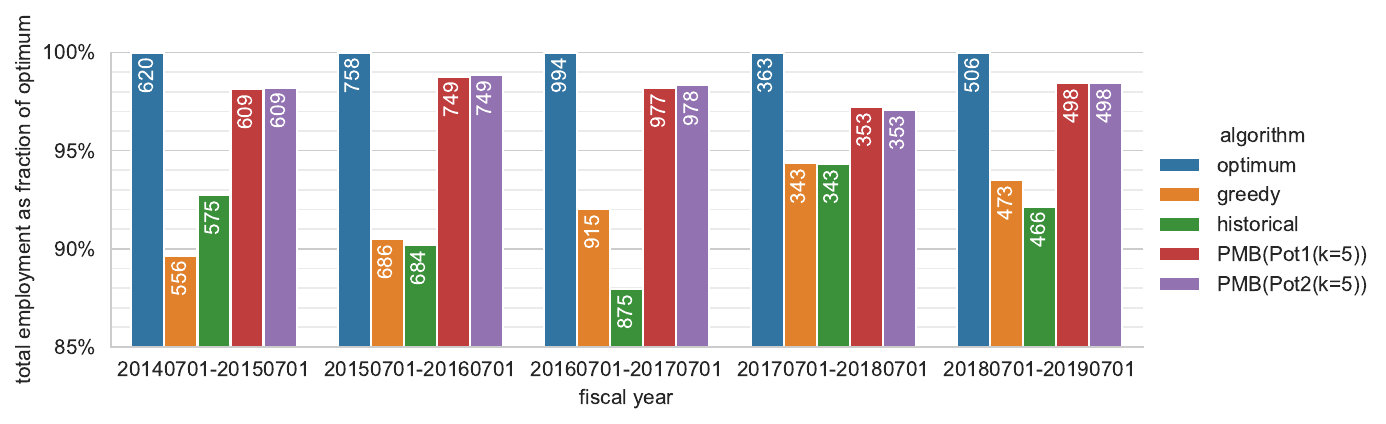}
    \caption{Total employment, where cases are not split and arrive in batches, but grouping arrivals into ``fiscal years'' lasting from April to March or from July to June rather than from October to September. Capacities are 110\% of the historical allocation of these cases. For the potential algorithms, total employment is averaged over 10 random runs.}
    \label{fig:april}
\end{figure}
Second, in \cref{fig:april}, we group historic arrivals not by which fiscal year (October to September) they arrive in, but make an assumption that the matching periods instead last from April to March or from July to June.
Since the capacities are not derived from any official fiscal-year capacities, but just set to 110\% of the historical arrivals during the matching period, the employment numbers of this experiment are not immediately comparable to the numbers displayed in other figures.
Again, the potential algorithms clearly outperform the historical and greedy matching.
In the matching periods April 2018--March 2019 and July 2017--June 2018, the employment of the potential algorithms slightly drops relative to what we saw in the main text's experiments, to around 97\% of the optimum.
In both of these periods, the optimal employment is particularly low, which most likely reflects low numbers of arriving cases and makes larger relative deviations more likely.
}

\subsection{Employment statistics if capacities are changed during the year}
\label{app:changingquotas}
For the fiscal years 2017 and 2018, we run a version of the experiments in \cref{sec:nfromcaps}, in which the capacities are actually reduced at the time they were revised.
For the greedy algorithm and the potential algorithms, this means that they are initially run using the initially announced capacities (and the potential algorithms expect 91\% of the initial total capacity to arrive).
At the time the capacity was revised, the capacity for the greedy algorithm and the potential algorithms is updated to the revised capacity.
If the revised capacity of an affiliate lies below what the algorithm already used before the point of revision, the capacity is frozen at current occupancy.
As we argue in the body of the paper, this gives the greedy algorithm an unfair advantage and makes the resulting total employment hard to compare.
The historical matching is roughly comparable since HIAS also started out aiming for the full capacities and later aimed for the revised capacities.
However, the historical matching in 2017 actually made use of a fairly large end-of-year increase in capacity, which the other algorithms do not have access to, and it profits from the fact that HIAS influenced the change in capacity to match its prior allocation decisions.
As a reference, we show optimal matchings in hindsight both for the initial and for the revised capacities.

\begin{figure}[H]
\centering
\includegraphics[width=.9\textwidth]{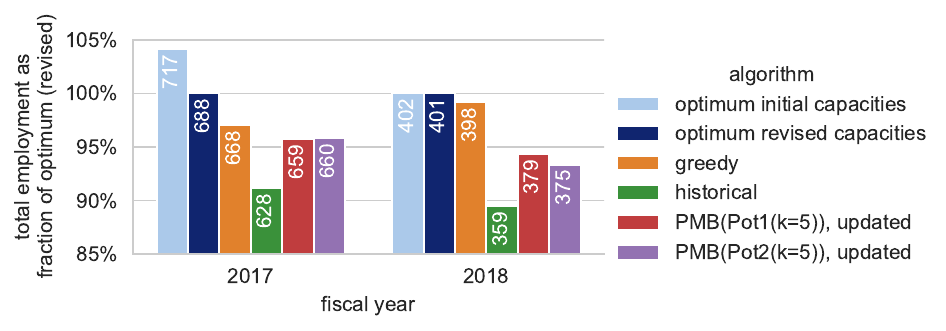}
\caption{Total employment, where cases are not split up and arrive in batches. Dynamic algorithms start respecting initial capacities, but capacities are changed at time of revision (except for historical). The potential algorithms do not have access to the true number of arriving cases but assume that the arriving refugees amount to 91\% of the initial, then the revised capacity. For potential algorithms, a single random run is shown.}
\label{fig:changing_caps}
\end{figure}

\subsection{Percentage of Matched Refugees}
\label{sec:unmatched}
Here, we report the percentage of refugees (i.e., cases weighted by case size) that gets matched to a real affiliate rather than the unmatched affiliate $\bot$.
In our implementation of \textsf{PMB} (\cref{alg:batching}), for a small constant $\epsilon$, we add a weight of $\epsilon \cdot s_i$ for all variables $x_{i, \ell}$ indicating that a case $i$ is matched to an affiliate $\ell \neq \bot$, with the intention of breaking ties between optimal solutions in favor of solutions that don't leave more cases unmatched than necessary.
This is particularly relevant since some cases, for example those consisting of unaccompanied minors, have an employment score of zero for all affiliates.
This added constant (and the equivalent modification to \textsf{PM}) ensures that whether these cases ever get matched does not depend on the implementation of the ILP solver.
Since we implement the greedy algorithm by setting potentials in \textsf{PMB} to zero, the same holds true for greedy.
Finally, a similar issue arises for the optimum matching in hindsight.
For this, we optimize total employment without further constraints on the number of matched refugees, but then greedily add cases with zero unemployment where they fit throughout the year.

In particular, this means that the optimum is not an upper bound on how many people are matched.
Instead, it is the historical matching that matches 100\% of refugees, but this comparison is not on equal terms since the historical matching conforms to final capacities that were specifically increased to fit all refugees and since it ignores some case--affiliate incompatibilities.

\begin{figure}[H]
    \centering
    \includegraphics[width=\textwidth]{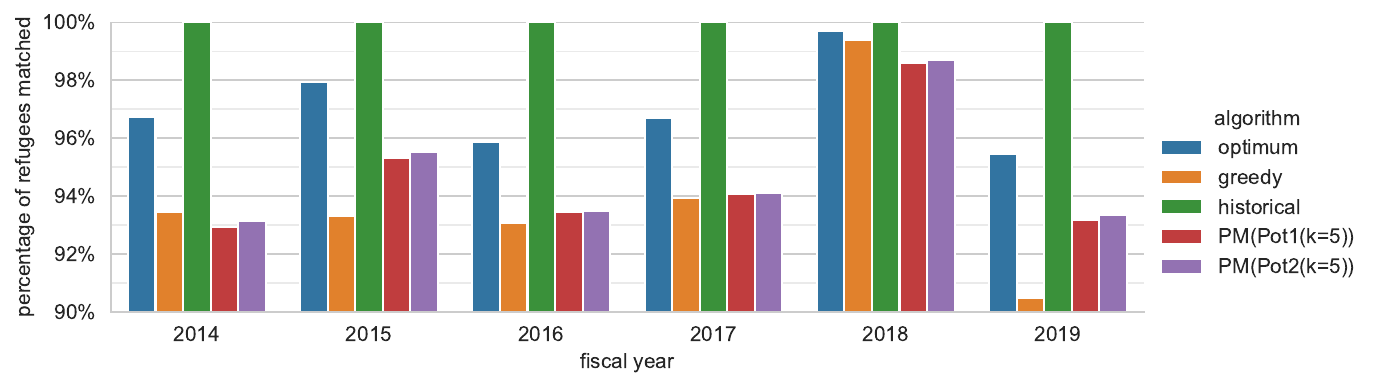}
    \caption{Match percentages for the experiment in \cref{fig:unit} (split cases, final capacities).}
    \label{fig:matched_unit}
\end{figure}

For the setting of online bipartite matching, we find in \cref{fig:matched_unit} that the potential algorithms are roughly on par with the greedy algorithm when it comes to matching many refugees.
The fact that both greedy and potential algorithms tend to leave a single-digit percentage of refugees unmatched can probably be explained by the many cases with family ties, which can only go in a single affiliate and have to remain unmatched if this affiliate is full, which is hard to avoid without knowing future arrivals.
The optimum-hindsight matching does not have this problem, and therefore matches around 2\% of refugees more on average.
The remaining gap to 100\% can be explained by the greedy matching of zero-employment cases, by the fact that the case--affiliate incompatibilities might not allow to match all cases (which the historical matching is not bound by), and by genuine trade-offs between matching more refugees and obtaining higher total employment.

\begin{figure}[H]
    \centering
    \includegraphics[width=\textwidth]{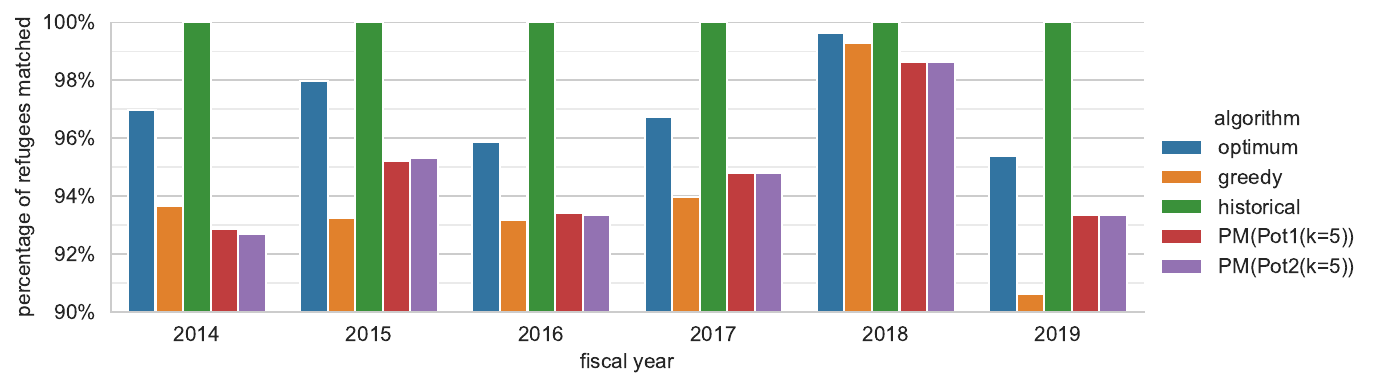}
    \caption{Match percentages for the experiment in \cref{fig:nobatching} (whole cases, final capacities).}
    \label{fig:matched_nobatching}
\end{figure}

\begin{figure}[H]
    \centering
    \includegraphics[width=\textwidth]{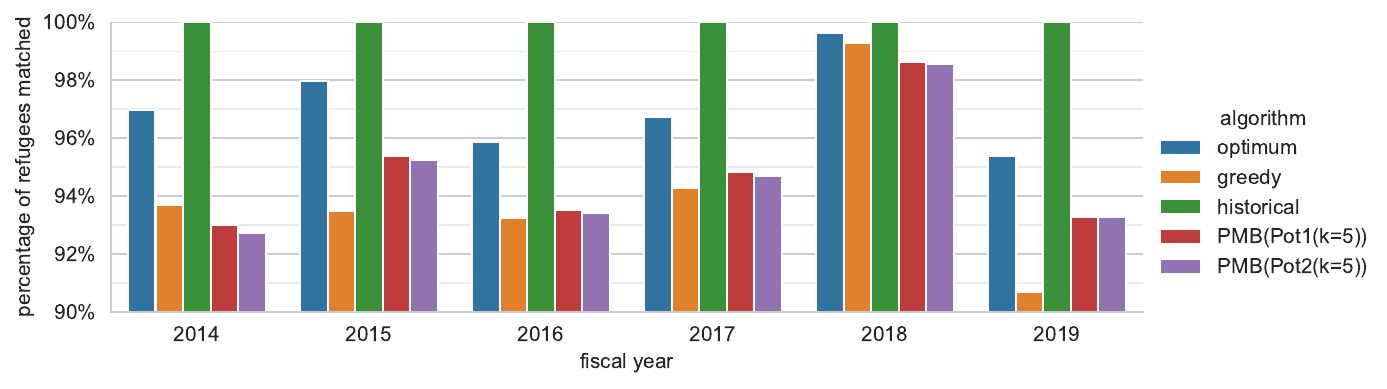}
    \caption{Match percentages for the experiment in \cref{fig:full} (whole cases, batching, final capacities).}
    \label{fig:matched_full}
\end{figure}

As for the total employment, the addition of non-unit case sizes (\cref{fig:matched_nobatching}) and of batching (\cref{fig:matched_full}) barely has any effect on the percentage of matched refugees.

\begin{figure}[H]
    \centering
    \includegraphics[width=\textwidth]{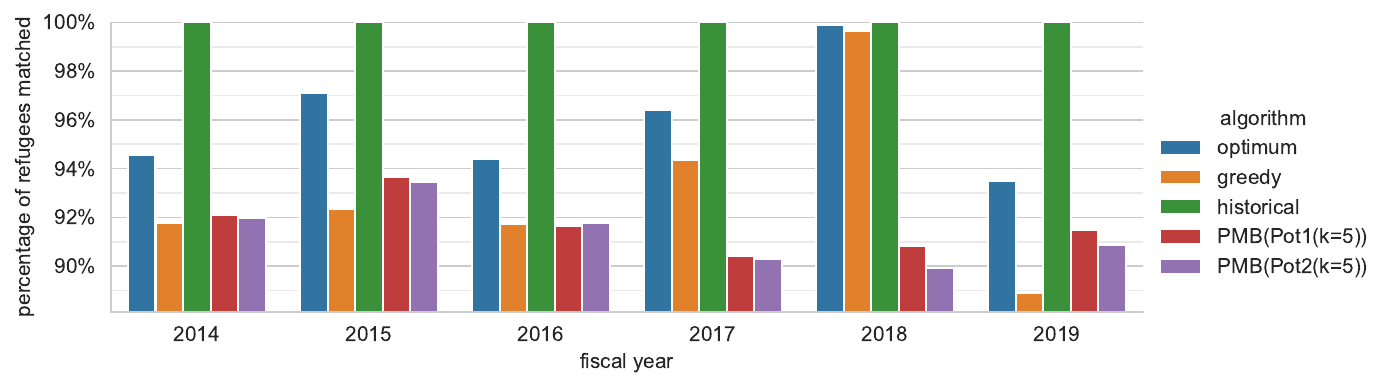}
    \caption{Match percentages for the experiment in \cref{fig:trustcaps} (whole cases, batches, initial capacities, potential algorithms do not know $n$).}
    \label{fig:matched_trustcaps}
\end{figure}

In \cref{fig:matched_trustcaps}, the percentage of matched refugees by the optimum matching and the greedy algorithm is lower in most years, which is due to the initial capacities being strictly tighter than the final capacities on all years other than 2017 and 2018.
The potential algorithms follow the same trend, but visibly match particularly few refugees in fiscal years 2017 and 2018.
Since these algorithms vastly overestimate the arrival numbers due to the overly large capacities, they leave some refugees unmatched in the expectation that this will benefit (spurious) later arrivals.

\end{document}